\documentstyle[12pt]{article}
\newcommand{\elltbar}{\overline{\widetilde{\ell}\, }}

\newcommand{\mysection}[1]{\setcounter{equation}{0}\section{#1}}

\newcommand{\mathbold}[1]{\mbox{\rm\bf #1}}
\newcommand{\mrm}[1]{\mbox{\rm #1}}

\newcommand{\bla}{\hspace{1cm}}
\newcommand{\beq}{\begin{equation}}
\newcommand{\bea}{\begin{eqnarray}}
\newcommand{\eea}{\end{eqnarray}}
\newcommand{\eq}[1]{eq.~(\ref{#1})}
\newcommand{\rfn}[1]{(\ref{#1})}

\newcommand{\db}{\hspace{-0.2ex}\not\hspace{-0.7ex}D\hspace{0.1ex}}
\newcommand{\sla}[1]{\hspace{-0.1ex}\not\hspace{-0.5ex} #1\hspace{0.1ex}}
\newcommand{\delte}{\Delta_\epsilon}
\newcommand{\abs}[1]{\left| #1\right|}
\newcommand{\Tr}[1]{\mathop{\mrm{Tr}}\left\{ #1 \right\}}
\newcommand{\tr}[1]{\mathop{\mrm{tr}}\left\{ #1 \right\}}
\newcommand{\vev}[1]{\left\langle #1\right\rangle}
\newcommand{\bra}[1]{\left\langle #1\right|}
\newcommand{\ket}[1]{\left| #1\right\rangle}
\newcommand{\lrover}[1]{
      \raisebox{1.3ex}{\rlap{$\leftrightarrow$}} \raisebox{ 0ex}{$#1$}}
\renewcommand{\titlepage}{\clearpage%
\setcounter{footnote}{0}%
\thispagestyle{empty}\pagestyle{plain}\pagenumbering{arabic}%
\kern1mm
\vskip15mm\normalsize}
\newcommand{\docnum}[1]{\hbox to \hsize{\hskip123mm\hbox{#1}\hss}}
\renewcommand{\date}[1]{\hbox to \hsize{\hskip123mm\hbox{#1}\hss}}

\renewcommand{\title}[1]{\vskip1em\begin{center}\Large\bf#1
\end{center}\vskip2.5em}
\renewcommand{\author}[1]{\vskip0.5em{\bf #1}\vskip0.5em}
\newcommand{\inst}[1]{\vskip0.3em{ #1}\vskip0.5em}
\renewcommand{\abstract}{\begin{center}{\bf Abstract}
\end{center}\quotation}

\newcommand{\anotfoot}[2]{\vfill\noindent\underline{\hspace{6cm}}
\par\noindent #1) #2}
\newcommand{\anotfootnb}[2]{\par\noindent #1) #2}

\begin{document}
\begin{titlepage}
\docnum{CERN-TH.7030/93}
\docnum{BI-TP 93/48}
\vspace{0.5cm}
\title{One-loop Effective Lagrangian for\\
a Standard Model with a Heavy Charged Scalar Singlet}
\begin{center}
\author{Mikhail Bilenky$^{*)}$}
\inst{Department of Physics\\
Univ.  Bielefeld, Postfach 8640\\
 D-33615  Bielefeld, Germany}
\inst{and}
\author{Arcadi Santamaria$^{**)}$}
\inst{TH Division, CERN, 1211 Gen\`eve 23, Switzerland}
\end{center}
\vspace{0.25cm}
\begin{abstract}
We study several problems related to the construction
and the use of effective Lagrangians by considering an
extension of the standard model that includes  a heavy  scalar singlet
 coupled to the leptonic doublet.
 Starting from the  full renormalizable model,
 we build an effective field theory  by integrating  out the heavy
 scalar.  A local effective Lagrangian (up to operators of dimension six)
 is obtained by expanding the
 one-loop effective action in inverse powers of the heavy mass.
 This is done by matching some
 Green functions calculated with both the full and the effective theories.

  Using this simple example we study the renormalization of  effective
 Lagrangians in general and discuss how they can be used to bound new physics.
 We also discuss the effective Lagrangian after spontaneous symmetry
 breaking,
 and the use of the standard model classical equations of motion to rewrite it
 in different forms. The final effective Lagrangian in the physical basis is
wel
 suited to the study of the phenomenology of the model, which we comment on
 briefly. Finally, as an example of the use of our effective field theory,
 we consider the leptonic flavour-changing decay
 of the $Z$ boson in the
 effective theory and compare the results obtained with the full model
calculation.
 \end{abstract}
 \vspace{0.25cm}
\vfill\noindent
CERN-TH.7030/93\\
October 1993
\anotfoot{*}{Alexander von Humboldt Fellow. On leave of absence from
the Joint Institute for Nuclear Research, Dubna, Russia.}
\anotfootnb{**}{On leave of absence from Departament de
F\'{\i}sica Te\`orica,
Universitat de Val\`encia, and IFIC, Val\`encia, Spain.}
\end{titlepage}
\setcounter{page}{1}

\mysection{Introduction}

Effective
Lagrangians
\cite{Wei67,Wei68,CWZ69,Zim70,Wei79,KY79,Wei80,OS80,KY82,GL84,GL85,Geo91}
have been used for a long time\footnote{Very early examples
are the effective Lagrangian description of the photon-photon interactions
\cite{EK35,Eul36,HE36} or Fermi's theory of weak
interactions \cite{Fer34,Fer34a,Fer34b}.}
as a systematic method to incorporate the known symmetries
of a problem into the quantum field theory language. However,  with the
advent of  Yang-Mills theories and the Higgs mechanism, which allowed the
construction of physically interesting renormalizable theories, they were
used only for those problems that could not be treated in any other way.
Particularly important have been its applications to low-energy strong
interactions in the form of the so-called  Chiral Perturbation Theory
\cite{Wei67,Wei68,CWZ69,Wei79,GL84,GL85,Geo84} and more recently
the  Heavy Quark Effective Field Theory \cite{IW89,IW90}.
Its application to weak interactions has been less intensive, since in the
last decade the main effort has been in the direction of building  complete
renormalizable theories that could serve as alternatives or extensions of the
standard model,  by just enlarging the number of fermions,  gauge bosons
and  scalars.  Renormalizability was considered to be a main point
and completely linked to predictability,  because non-renormalizable models
need an infinite number of parameters to be completely described.
The striking confirmation of many of the  standard model
predictions  in LEP experiments has started to change
this point of view.  Almost every one now thinks that the
standard model correctly describes physics at present energies and perhaps
 also up to energies close to the TeV range.
 Of course, supersymmetric
particles and other elusive particles,  such as neutral heavy  leptons or
right-handed neutrinos,  with some hidden interactions and
masses much lighter than 1 TeV, are not excluded.
On the other hand, the theoretical developments achieved since the
beginning of
the seventies have brought a new interpretation of renormalizability
\cite{Geo84,Wil71,WK74,Pol84,Pol92,Wei92}. Nowadays
renormalizability is not understood as a calculational requirement or
a consistency requirement. In fact one knows how to calculate, for example,
with the Fermi Lagrangian, as long as one does not try to use it
beyond  the Fermi scale.  The key idea is that one does not
expect a  quantum field theory,  even a renormalizable one,
to be valid up to
arbitrarily large energy scales. Renormalizability is then seen as the physical
requirement that physics at low energies cannot dramatically depend
on the physics at some large scale.  There could be effects of the heavy
particles on the low-energy physics,  but all of them must be
suppressed by
some power of the scale $\Lambda$ at which the new physics starts.
Then, for each range
of energies one expects that physics is described by a Lagrangian of the
form:
\begin{equation}
\label{firsteq}
{\cal L}_{eff}= {\cal L}_0 + \frac{1}{\Lambda} {\cal L}_1 +
                        \frac{1}{\Lambda^2}{\cal L}_2+ \cdots\ ,
\end{equation}
 where ${\cal L}_0$ is a renormalizable Lagrangian that describes
 the low-energy physics and ${\cal L}_n$ are linear combinations of
 non-renormalizable operators of dimension $n+4$.  For processes involving
 energies $E$ much smaller than $\Lambda$, the effects of the
non-renormalizable
 operators are suppressed by  $(E/\Lambda)^n$  and can systematically be
 computed. Of course, they depend on more arbitrary parameters, but, if
 the underlying theory is known and if it is renormalizable, they can,  in
 principle,  be computed in terms of the few renormalizable couplings of
 the underlying theory by matching the calculation of several observables
 or Green functions done with  both theories: the full theory and the effective
 one.  This picture is known as the decoupling case \cite{AC75},
 as the effects of
 heavy particles decouple from the low-energy physics, and a
 well-known example is physics in the 5--80~GeV range. There,  physics can
 be well described by a $QED$-$QCD$ renormalizable gauge-invariant
 Lagrangian involving the five light quarks and the leptons, supplemented
 by four-fermion  non-renormalizable interactions that take
 into account  the effects of weak interactions.  These
 non-renormalizable couplings can be computed in terms of the standard
 model couplings by matching some Green functions at the Fermi scale.
 But once this matching is done all calculations, including $QED$ and $QCD$
 radiative corrections, can be performed at the effective Lagrangian level.
 Since the full standard  model is $QED$ and $QCD$ gauge invariant one
 obtains that all the effective couplings must also be $QED$ and $QCD$
 gauge invariant.

 Following the previous reasoning
 one can then  expect that the standard model is not valid for an
 arbitrarily
 large range of energies and that it is just a low-energy approximation
 of a more complete theory. There are many arguments suggesting
 that this is the case. In particular, the so-called naturalness
 problem of the standard model, which will be discussed later on in the
context of
 our effective Lagrangian, and the large amount of arbitrary input
 parameters needed to describe the standard model are among the most
popular
 ones. If the standard model is just a low-energy approximation of a more
complete
 theory,  one would expect that new
 non-renormalizable couplings suppressed by the scale of the new
 physics should arise.  In complete parallelism one would also expect
 these non-renormalizable interactions to be gauge invariant with respect
 to the standard model gauge group.  However, a complication arises since
 the $SU(2)\otimes U(1)$ gauge symmetry of the standard model is
 spontaneously broken to $U(1)_{QED}$. Then, when building an effective
 theory for physics beyond the standard model, there are two possibilities:
 \begin{itemize}
\item  The gauge symmetry is linearly realized. This means that radial
excitations of the scalar field are relatively light; then, for energies
larger than the mass of these excitations, the symmetry is effectively
restored. This is the simplest situation,
in which the effective Lagrangian is constructed with exactly the same
particle content as the standard model particle spectrum
and non-renormalizable
interactions are required to be $SU(2)\otimes U(1)$ gauge invariant.
\item  The gauge symmetry is non-linearly realized. That is,  the physics of
spontaneous symmetry breaking is non-linear. The radial excitations
are heavy with respect to the Fermi scale; then, there is a wide range
of energies between the Fermi scale and the scale of  new physics in which
the full standard gauge symmetry is present, but realized in a non-linear
way.  In this region of energies one can still write an effective
Lagrangian, but the first term of that Lagrangian is a non-renormalizable
non-linear sigma model \cite{AB80,Lon80,Lon81,Che88,DE91,EH92}.

\end{itemize}
 There is a big difference between these two possibilities. The second one
 assumes that the new physics  is responsible for spontaneous symmetry
 breaking; as a consequence it must start not far away from the Fermi
 scale,
 since it
 must correct the bad behaviour of the standard model without Higgs
 particles.  In  the first one, however, the Higgs mechanism is fully
 implemented.  Low-energy physics decouples completely from
 high-energy physics and the only way to get some hint on the scale of
 the possible new physics (apart from naturalness arguments) is
 just by looking at the size of the non-renormalizable
 interactions. Experimental bounds are then the only source of
 information.

 A very simple and instructive illustration of the use of effective
 Lagrangians to bound new physics
 is the so-called see-saw mechanism for the generation of neutrino
 masses \cite{GRS80,Yan79}. It is not difficult to see that the only
 $SU(2)\otimes U(1)$
 gauge-invariant operator of dimension five that can be built with the field
 content of the standard model is
 \begin{equation}
 \label{see-saw1}
  {\cal L}_{see-saw}  =  -\frac{1}{4} \frac{1}{\Lambda}
  (\elltbar F \vec{\tau} \ell)
  (\widetilde{\varphi}^\dagger \vec{\tau} \varphi) \ ,
 \end{equation}
where $\ell$ is the standard left-handed doublet of leptons,
$\widetilde{\ell} = i\tau_2 \ell^c$,  $\ell^c =
C\overline{\ell}^T$
( $C$ is the charge conjugation operator),
$\varphi$ is
the Higgs doublet and $\widetilde{\varphi} = i\tau_2 \varphi^*$; $F$ is a
complex symmetric matrix in flavour space ($SU(2)$ and flavour indices have
been suppressed). It is clear that this Lagrangian does not conserve
generational lepton numbers,  but in addition it does not conserve the
total lepton
number,  which is violated in two units.  This kind of  operator will be
generated in any theory that does not conserve lepton number.
When the Higgs develops a vacuum expectation value (VEV),
it will generate a neutrino Majorana mass matrix given
by
\begin{equation}
\label{mnu}
 M_{\nu} = F \frac{\vev{\varphi^{(0)}}^2}{\Lambda} \ .
\end{equation}
If we take the largest eigenvalue of  $F$ to be of order 1,
the Higgs VEV $\vev{\varphi^{(0)}}$=~174~GeV and
use the experimental bound on the
$\tau$-neutrino mass, $m_{\nu_{\tau}} < 31$~MeV, we find that  $\Lambda  >
10^6$~GeV.  Should one take the cosmological bound \cite{CM72}
on neutrino masses, $m_{\nu_{\tau}} < 100$~eV, one would obtain
$\Lambda > 3\times 10^{11}$~GeV. These bounds are really impressive.
But what is $\Lambda$?
The Lagrangian of \eq{see-saw1} seems to be generated by the exchange of a
scalar triplet with hypercharge 1 between the leptons and the Higgses.  Then,
$\Lambda$ should be the mass of that triplet. However
this is not the only possibility. In fact, \eq{see-saw1} can be identically
rewritten,
after a $SU(2)$ Fierz transformation, as
\begin{equation}
 \label{see-saw2}
  {\cal L}_{see-saw}  =  - \frac{1}{2}\frac{1}{\Lambda}
   (\elltbar \varphi) F  (\widetilde{\varphi}^\dagger \ell)
 \ ,
 \end{equation}
 which suggests the exchange of a neutral heavy Majorana fermion; then
 $\Lambda$ should be the mass of that fermion. Indeed, the original
 formulation of the see-saw mechanism\cite{GRS80,Yan79} was based on
 this possibility.

 This simple example shows the power and the limitations  of the effective
 Lagrangian approach. One can set impressive bounds on the scale of new
 physics, but one cannot  completely disentangle its origin,  at least by
taking
 into account only the lowest-order  operators.

 It must also  be
 remarked that in the effective Lagrangian approach it is essential
 to know what  the low-energy
 particle spectrum is. One generally assumes that it is just the standard
 model particle spectrum, but there could exist
 light particles, completely neutral under the standard model gauge group,
 that interact with the standard model particles only through the exchange of
 new heavy particles.
 This is not at all an exotic situation since it is what happened with
 weak interactions: neutrinos are singlets under $QCD$ and $QED$; however,
 they cannot be ignored in building an effective Lagrangian that describes
 weak interactions.

 Keeping in mind all these limitations, one can very efficiently use
 the effective
 Lagrangian approach to new physics
 to set bounds on  operators that violate some of the global symmetries
 of the standard
 model:  lepton number conservation,  baryon number conservation or
 flavour symmetries \cite{Wei79a,WZ79,WZ80,BW86}.
 These bounds  are naturally implemented at tree level in the
 effective Lagrangian (although the effective operators could be generated
 through loops in the full theory).  During the last years, using
 the  precise data obtained from experiment,
 people have started to consider the possibility of
 bounding  operators that do not violate any symmetry of the standard
 model \cite{BW86,GW91}. This task
 is much more complicated because the number of
 operators of a given dimension is very large \cite{BW86,BS83,LLR86}
 (after using the equations
 of motion and without taking into account flavour, Buchm\"uller and Wyler
 \cite{BW86}
 found 80 $SU(2)\otimes U(1)$ gauge invariant dimension-six operators
 constructed from the standard model fields). Especially interesting is the
 analysis
 of effective operators contributing to trilinear gauge-boson couplings,
 since they could have important
 consequences at LEP2,  the SSC and LHC \cite{BKR92,GR93,DGH92}.
 Some of these operators also contribute to
 LEP1 observables at tree level and they are strongly bounded by the LEP
 very precise measurements. Others do not contribute at tree level to
 LEP1 observables, but only in loops \cite{DGH92,HV93,HIS92,HIS93}.
  The next step is to use the effective Lagrangian at the one-loop level and
 try to set bounds on all operators by using radiative corrections.  This
 analysis is even more complicated since the effective theory is
 non-renormalizable in the standard sense and care must be taken
 with divergences,
 especially when the full theory is not known. If the full theory is known,
 it is not difficult to find the right prescription to absorb all
 infinities in the effective theory.
 All couplings must be renormalized, and matching to the full
 theory fixes all the
 counterterms. If the full theory is not known, the theory still has to be
 renormalized  and infinities absorbed in the various couplings; however,
 the finite parts of the counterterms remain arbitrary and cannot be
 determined. The way this is done has created some controversy in the
 literature. The most general trilinear interaction among vector bosons
 can be constructed by imposing only Lorentz invariance and QED gauge
 invariance \cite{GG79,HPZ87}.
 In the last years several groups have used  such interactions in loops.
  Calculations were performed by using a momentum cut-off that was
 identified with the scale of new physics. Then, large effects were found
 since, in some cases, the diagrams were quadratically or even quartically
 divergent\footnote{
 A long list of references in which
 this method was used can be found in \cite{DGH92,BL92a}.}. This
 approach is not always correct, because
 the effective theory also has to be renormalized,
 couplings must
 be defined at some scale and they must  satisfy some renormalization group
 equations. Criticisms to this treatment of effective Lagrangians have already
 been raised by several groups
 \cite{DGH92,HV93,HIS92,HIS93,BL92a,BL92,AEW92,Ein93,Wud93}; however,
  the emphasis was put on different
 aspects.  The authors of refs.~\cite{DGH92,HV93} stressed the importance
 of  the gauge invariance of the effective Lagrangian under the standard
 model gauge group. While, for example, in refs.~\cite{BL92a,BL92}
 the emphasis
 was put on the incorrect use of cut-offs in previous calculations.

 The purpose of this paper is to study some of the questions that
 arise when the effective Lagrangian approach to new physics is used
 by working out completely
  an example of possible new physics. We consider the
  simplest non-trivial extension of
 the standard model we could write down: a standard model supplemented
 by a singly charged scalar singlet coupled to leptons \cite{CL80,Zee80}.
 We construct the full model and obtain a low-energy effective field
 theory by integrating out, at the one-loop level,  the heavy scalar
 singlet.
 By construction, since the full theory is $SU(2)\otimes U(1)$ gauge
 invariant and we integrate out a complete scalar multiplet,
 we automatically obtain a $SU(2)\otimes U(1)$ gauge invariant
 effective theory.  Therefore,  we do not  discuss the role of
 gauge invariance in the construction of the effective theory
\footnote{It seems,
 however, that if the full theory is gauge invariant with respect to some
 group, the effective theory should also be gauge invariant,
 although gauge invariance
  could be implemented in a non-linear way.
 The effective
 theories obtained by integrating out the standard Higgs
 \cite{AB80,Lon80,Lon81,Che88,EH92} or a heavy  quark
 \cite{DF84,DF84a,SFY87,LSY91,LSY93,FMM92}
 in the standard model belong to this second type of theories.}.

  We use this example to study
 the renormalization of the effective Lagrangian and to study the
 matching conditions that relate the parameters of the effective Lagrangian
  to the parameters of the full Lagrangian ensuring agreement between
 the two theories at low energies. The possibility of using the equations of
 motion before or after spontaneous symmetry breaking is illustrated as well.
 The model is  also phenomenologically interesting because the presence
 of the scalar gives rise to very interesting
 phenomena \cite{CL80,Zee80,Pet82,LTV85,BS88} such as, for example,
 the processes $Z \rightarrow \tau e$, $Z \rightarrow \mu e$, $\cdots$,
  $\mu \rightarrow e \gamma$, $\tau \rightarrow \mu \gamma$,  $\cdots$,
  $\mu \rightarrow e e e$, $\tau \rightarrow \mu e e$, $\cdots$. There
  are also additional contributions to the masses of the gauge bosons,
  new neutral current interactions, etc.

   Of course, this analysis is quite far from being general, it is a
  very specific model that leads to a linearly realized gauge symmetry,
  but in spite of its simplicity it leads, already at the one-loop level,
  to many of the operators classified in refs.
  \cite{BW86,BS83,LLR86}.
   It can also help us to understand some of the tricky points discussed in the
  literature.

  In section
 \ref{fulllagrangian} we will introduce the notation and write down the
 Lagrangian of the full theory.  In section \ref{intscalar} we obtain the
 tree-level and the one-loop contributions to the effective Lagrangian
 of diagrams with only heavy scalars in internal lines by using functional
 methods.
 When using the tree-level effective interactions at the one-loop
 level in the effective Lagrangian, new effective operators appear as
 counterterms that must be computed by matching the full theory results
 with the effective Lagrangian results; we do this job in section
 \ref{matching}. Renormalization of our effective Lagrangian
 and the possibility of using effective Lagrangians in general at
 the one-loop level to bound new physics are discussed in section
 \ref{renormalization}.
 In section \ref{ssb} we discuss  the use of the classical equations of motion
 to rewrite the effective Lagrangian in a convenient form for
 phenomenological analyses. We also discuss the impact of
 spontaneous symmetry breaking on this Lagrangian and
 extract the relevant interactions in terms of the physical fields.
  In section \ref{pheno} we shortly
 comment on
some of the most
 interesting phenomenological consequences of the model by using the
 effective Lagrangian we have obtained. Finally, in section
 \ref{conclusion} we review what we have learned about the use of
 effective Lagrangians with our example.
  We include in  appendix \ref{det} the calculation of the
 determinant of the fluctuation operator,  in
 appendix~\ref{heavy-light} the calculation of the diagrams with
 heavy-light lines,
 and finally in  appendix \ref{fcdecay} we calculate the amplitudes for
 $Z\rightarrow \bar{e}_a e_b$ in both the full and the effective theory.

\mysection{The Model}
\label{fulllagrangian}

The complete renormalizable model we are considering is
an extension of the standard model, which contains
a singly charged scalar singlet in addition to the standard model
particles.  This model is one of the simplest extensions one
could imagine, but in spite of its simplicity, it already includes many
interesting features common to any extension of the standard model
containing a large mass scale compared  with  the Fermi scale.
For completeness we list below the particle spectrum of the full
model and give the corresponding transformation properties
with respect to the gauge group $SU(3)\otimes SU(2) \otimes U(1)$\\
\vskip 0.5cm
\begin{tabular}{lll}
left-handed lepton doublets: & $\ell$  &   $(1,~2,-1/2)$\\
& $\widetilde{\ell} = i\tau_2 \ell^c$ & $(1,~2,~~~1/2)$\\
right-handed charged leptons: & $e$  & $(1,~1,-1~~~~)$\\
left-handed quark doublets: & $q$  & $(3,~2,~~~1/6)$\\
right-handed $u$-quarks: &$u$   & $(3,~1,~~~2/3)$\\
right-handed $d$-quarks: &$d$  &  $(3,~1,-1/3)$\\
Higgs boson doublet: & $\varphi$ & $(1,~2,~~~1/2)$\\
      & $\widetilde{\varphi} = i\tau_2 \varphi^*$ & $(1,~2,-1/2)$\\
gluons:  & $G^a_\mu$  &  $(8,~1,~~~0~~~~)$\\
     & $G^a_{\mu\nu} =\partial_\mu G^a_\nu-\partial_\nu G^a_\mu+
         g_s f^a_{\ bc} G^b_\mu G^c_\nu$ & \\
$W$ bosons:  & $\vec{W}_\mu$ &   $(1,~3,~~~0~~~~)$\\
         & $\vec{W}_{\mu\nu} =
         \partial_\mu \vec{W}_\nu-\partial_\nu \vec{W}_\mu+
         g \vec{W}_\mu \times \vec{W}_\nu$ & \\
$B$ bosons:   & $B_\mu$  & $(1,~1,~~~0~~~~)$ \\
                    & $B_{\mu\nu}=
                    \partial_\mu B_\nu-\partial_\nu B_\mu$  & \\
charged scalar singlet:  & $h\ $  &  $(1,~1,-1~~~~)$\ .\\
\end{tabular}
\vskip 0.5cm
\noindent  All possible $SU(2)$ and generational indices are
suppressed above.

The full Lagrangian  can be split
into two parts:
\begin{equation}
\label{lagrangian}
{\cal L}_{full} = {\cal L}_{SM} + {\cal L}_h\  .
\end{equation}
The first part, ${\cal L}_{SM}$,
 represents the standard model Lagrangian:
\begin{eqnarray}
{\cal L}_{SM} =  -\frac{1}{4} G^a_{\mu\nu} G_a^{\mu\nu}
-\frac{1}{4} \vec{W}_{\mu\nu} \vec{W}^{\mu\nu}
-\frac{1}{4} B_{\mu\nu} B^{\mu\nu}
  + (D_\mu \varphi)^\dagger (D^\mu \varphi) + m_{\varphi}^2
\varphi^\dagger\varphi - \lambda (\varphi^\dagger\varphi)^2  \nonumber\\
  +i \overline{\ell} \db \ell + i \overline{e} \db e + i \overline{q} \db q
+i \overline{u} \db u +i \overline{d} \db d
  + ( \overline{\ell} Y_e e \varphi + \overline{q} Y_d d \varphi+
\overline{q} Y_u u \widetilde{\varphi} + \mrm{h.c.})\ ,
\label{smlagrangian}
\end{eqnarray}
where the covariant derivative can be written in the general form
\begin{equation}
\label{covderiv}
D_\mu = \partial_\mu -i g_s T_3^a G^a_\mu
-i g T_2^i W^i_\mu -i g' Y B_\mu \ .
\end{equation}
 By $T_2$ and $T_3$ we denote
the generators of the
$SU(2)$ and the $SU(3)$ groups, accordingly, which are
acting on the
proper representations of the groups. In the case of $SU(2)$ doublets
$T_2^i=\frac{1}{2}\tau^i$, where $\tau^i$ are $2 \times 2$ Pauli
matrices. For
$SU(3)$-triplets $T_3^a= \frac{1}{2}\lambda^a$, where $\lambda^a$
are $3 \times 3$ Gell-Mann matrices.
In the Lagrangian \rfn{smlagrangian}
the Yukawa couplings, $Y_e, Y_d, Y_u$, are arbitrary
complex matrices in  flavour space.

The Lagrangian  ${\cal L}_h$ contains the  interactions
of the scalar singlet:
\begin{equation}
\label{lsinglet}
{\cal L}_h=
 (D_\mu  h )^\dagger D^\mu h
- m^2 \abs{h}^2 - \alpha \abs{h}^4
- \beta \abs{h}^2 \varphi^\dagger\varphi
+\left( \elltbar f \ell h^+ + \mrm{h.c.}\right)
\end{equation}
and the covariant derivative in \eq{lsinglet}
has the form  $D_\mu = \partial_\mu +i g' B_\mu$,  because
the $h$-scalar is an $SU(2)$ singlet with hypercharge $Y=-1$.

For further applications
it is convenient to rewrite the $h$-dependent part of the Lagrangian,
using integration by parts for the covariant derivative,
 in the following form:
\begin{equation}
\label{hlag}
{\cal L}_h = h^+ (-D^2-m^2-\alpha \abs{h}^2-\beta \varphi^\dagger
\varphi) h +\left( \elltbar f \ell h^+ + \mrm{h.c.}\right)~.
\end{equation}

The quantum numbers of the scalar singlet are such that
there is no interaction between the scalar singlet $h$ and the quarks in
the Lagrangian of \eq{lsinglet}.
It is also important to note that its  coupling
to the leptons, $f$, is
an antisymmetric complex matrix in flavour space.
The antisymmetry of this matrix
is a consequence of Fermi statistics and the fact that the current
$ \elltbar_a \ell_b$ is an $SU(2)$
scalar.
Then, one can write
$ \elltbar_a \ell_b =- \elltbar_b \ell_a$,
from which the antisymmetry of the coupling easily follows.
 This property of the scalar-lepton coupling is very important
since it naturally leads to a violation of  the generational
lepton numbers and all the interesting phenomenology related
to it.
However, if every lepton carries a total lepton number of 1,  we can
assign a total lepton number of 2 to the scalar $h$ and, then, the
Lagrangian \rfn{lsinglet} is invariant with respect to a global symmetry,
which can be identified as conservation of the total lepton number.
As a consequence, the neutrinos remain massless at all orders.
Models very similar to the one defined in \eq{lagrangian} containing
additional doublets and/or scalar singlets, in which the total
lepton number
is explicitly or spontaneously broken,
have been used  \cite{CL80,Zee80,LTV85,BS88,AR91,BS91} to
generate small calculable neutrino masses.

\mysection{Integrating out the heavy scalar}
\label{intscalar}

If the mass $m$ of the scalar $h$ is much larger than the
energy available to experiment,  the only
effects of this particle on the low-energy observables
come through virtual corrections. These effects can be taken
into account by using an effective Lagrangian of the form \rfn{firsteq},
containing only the ``light'' fields of the model and where
the effects of the ``heavy'' particle are included in the
non-renormalizable terms of dimension  larger than four,
which are suppressed by inverse
powers of the heavy mass $m$.

The effective Lagrangian can be defined
 by integrating out the heavy scalar field
 \cite{Wei80,Wil71,WK74,Pol84,Pol92} :
\begin{eqnarray}
\label{integraout}
e^{i S_{eff}}=
\exp\left\{ i \int d^4 x {\cal L}_{eff}(x) \right\} \equiv
\int {\cal D} h{\cal D} h^+  e^{i S} =
\int {\cal D} h{\cal D} h^+ \exp\left\{i \int d^4 x {\cal L}(x)\right\}
\\ \nonumber
= \exp\left\{ i\int d^4 x {\cal L}_{SM}(x) \right\}
\int {\cal D} h{\cal D} h^+ \exp\left\{i \int d^4 x
{\cal L}_h(x)\right\}
=e^{i S_{SM}}\int {\cal D} h{\cal D} h^+  e^{i S_h} ~,
\end{eqnarray}
where ${\cal D} h$ represents  functional integration over $h$.
The effective field theory defined by this  effective Lagrangian is fully
equivalent to the original theory when only Green functions with light external
particles are considered.
Starting from  \eq{integraout}
 we can calculate the one-loop level effective Lagrangian
by using the steepest-descent
method to integrate out the heavy scalar.
As we are interested in the effects of a heavy scalar
($m \geq 1$~TeV) on the physics around the
$M_Z$ scale and below, we will systematically
keep only terms of order $ O(1/m^2)$ through the whole calculation,
neglecting all operators with higher inverse powers of the mass of the
scalar singlet.

Let us denote by $h_0$ the solution of the classical equation of
motion for the $h$-field, i.e.
\begin{equation}
\left.\frac{\delta S}{\delta h(x) } \right|_{h_0} =0~.
\end{equation}
Using  the Lagrangian \rfn{hlag} we obtain :
\begin{equation}
\label{eqofmotion}
(-D^2-m^2-2\alpha \abs{h_0}^2-\beta \varphi^\dagger \varphi) h_0 +
\elltbar f \ell=0\ .
\end{equation}
Then the full action can be functionally expanded around the
solution $h_0(x)$
\begin{equation}
\label{expansion}
S= S_{SM} + S_h[h_0]
+ \int d^4x d^4 x'  \eta^\dagger(x) O(x,x') \eta(x') +\cdots~,
\end{equation}
where
the fluctuation operator $O(x,x')$ is given by the
second-order term in Taylor's expansion of the action $S$:
\begin{equation}
\label{fluct}
O(x,x') = \left.\frac{\delta^2 S}{\delta h(x) \delta h(x')} \right|_{h_0}
\end{equation}
and $\eta(x)$ in \eq{expansion}
represents the fluctuations around the classical
solution, $\eta(x)= h(x)-h_0(x)$.
 Substituting \eq{expansion} in \eq{integraout}, shifting variables
from $h$ to $\eta$ in the functional integration, and using the known
formula for Gaussian integration, we obtain
\begin{equation}
e^{iS_{eff}} = \exp\{i(S_{SM}+S_h[h_0])\} \det(O)^{-1} =
\exp \left\{i(S_{SM}+ S[h_0])-\Tr{\log(O)}\right\}~.
\end{equation}
Therefore, for the one-loop
effective action  we have
\begin{equation}
\label{seffgen}
S_{eff}= S_{SM}+S_h[h_0]+i \Tr{\log(O)}~.
 \end{equation}
 The first term in \eq{seffgen} is just the pure standard model
contribution to the effective action.
Using the equation of motion, \eq{eqofmotion}, the second term, $S_h[h_0]$,
can be
formally written as
\begin{equation}
\label{lnonlocal}
S_h[h_0] \approx
 -\int d^4x \elltbar(x) f \ell(x)
  \frac{1}{(-D^2-m^2-\beta \varphi^\dagger(x) \varphi(x))}
            \overline{\ell}(x) f^\dagger \widetilde{\ell}(x)\ ,
\end{equation}
where terms of order $1/m^4$ have been neglected.
The Lagrangian \rfn{lnonlocal} is highly non-local. To obtain a local
version of it we could further expand the operator\newline
${1}/{(-D^2-m^2-\beta \varphi^\dagger(x) \varphi(x))}$
as a power series in $1/m^2$:
\begin{equation}
\label{appsolution}
\frac{1}{(-D^2-m^2-\beta \varphi^\dagger(x) \varphi(x))}
= -\frac{1}{m^2}
+ \frac{1}{m^4} (D^2+\beta \varphi^\dagger(x) \varphi(x))+\cdots~.
\end{equation}
Keeping  the first term we obtain the only tree-level contribution,
at order $1/m^2$, of the scalar to the effective Lagrangian:
\begin{equation}
S_h[h_0]= \int d^4 x  {\cal L}^{(0)}(x)
\end{equation}
with
\begin{equation}
\label{lagr0}
{\cal L}^{(0)} =  \frac{1}{m^2}
 (\elltbar f\ell)
 (\overline{\ell}  f^\dagger\widetilde{\ell})=
 \frac{4}{m^2} f_{ab}  f^*_{a'b'}
 (\overline{\nu^c_{aL}} e_{bL})(\overline{e_{b'L}} \nu^c_{a'L})\ ,
\end{equation}
where the summation over repeated flavour indices
($a,b,a',b'$) is assumed.
It could also be obtained  by computing the tree-level
 diagram of fig.~1.a in the limit $q^2 \ll m^2$ (where $q$ is the momentum
 of the scalar).
The result in \eq{lagr0} is not complete if we are going to use this
Lagrangian for calculations at the loop level \cite{Wit76,Wit77}.
The reason is that  in order to obtain a
local Lagrangian we had to use the expansion \rfn{appsolution}.
 In doing this we assumed that the derivative
in \eq{appsolution}, or equivalently the momentum of the scalar,
is negligible  in comparison with the mass of the scalar.  But this is not true
if the $h$-scalar contributes
inside loops, since in  this case its momentum runs up to infinity.
It has been shown \cite{Wit76,Wit77} that the procedure just outlined
can be justified as long as one includes in the effective Lagrangian
a set of local operators that compensate for the terms missed
by using expansion \rfn{appsolution}.
We are going to compute these operators in the next section.

The last piece of the effective action \rfn{seffgen} is defined
in terms of  the fluctuation operator \rfn{fluct} and takes into account
all one-loop effects with only the heavy particle in loops. In our
approximation, keeping only terms $O(1/m^2)$, the
fluctuation operator \rfn{fluct} can be easily calculated and leads
to the following effective action
\begin{equation}
i \int d^4 x {\cal L}^{(1)}(x)=
\log{\det{(O)}^{-1}}=
-\Tr{\log(O)}=
-\Tr{\log(-D^2-m^2-\beta \varphi^\dagger \varphi)}\ .
\end{equation}

We moved to appendix~\ref{det} the detailed
evaluation of the determinant
of a generic operator of this form as an expansion in $1/m^2$.
Using the general result of  appendix~\ref{det}, \eq{resdet}, for our
 particular
case, $U=\beta \varphi^\dagger \varphi$,
$D_\mu =  \partial_\mu + i g' B_\mu$, that is $A_\mu \rightarrow i g' B_\mu$
and $F_{\mu\nu} \rightarrow  i g' B_{\mu\nu}$,
and taking into account that $D^\mu
 (\varphi^\dagger \varphi)=\partial^\mu (\varphi^\dagger \varphi)$,
 $D^\mu B_{\mu\nu}=\partial^\mu B_{\mu\nu}$,
we write this contribution to the effective Lagrangian
 as a sum of two parts:
\begin{equation}
\label{lfluctop}
{\cal L}^{(1)} = {\cal L}_{ren}^{(1)} + {\cal L}_{det}^{(1)} \  .
\end{equation}
 The first part, ${\cal L}_{ren}^{(1)}$,
contains only operators with dimension not larger than four. It
 is given by the following expression:
 \begin{equation}
\label{lrenorm}
{\cal L}_{ren}^{(1)} = \frac{1}{(4\pi)^2}\left(
\frac{m^4}{2}\left(\frac{3}{2}+\delte\right)+
m^2 \beta (1+\delte) (\varphi^\dagger\varphi)+
\frac{\beta^2}{2} \delte (\varphi^\dagger\varphi)^2
-\frac{g'^2}{12} \delte B_{\mu \nu} B^{\mu \nu}\right)~.
\end{equation}
All the
coefficients in the above Lagrangian are ultraviolet-divergent.
Here and in the rest of the paper we use dimensional regularization,
where these divergences appear as simple poles, $1/\epsilon$,
($\epsilon=2-D/2$) in
the function $\Delta_{\epsilon}$ defined in \eq{deltae} of
appendix \ref{det}.
As expected, all terms in \eq{lrenorm}  are already
present in the standard model Lagrangian. Therefore, they
can be absorbed in  a redefinition  of the parameters of the standard model.

In the second part of \eq{lfluctop}, ${\cal L}_{det}^{(1)}$, we included all
dimension-six operators:
\begin{eqnarray}
{\cal L}_{det}^{(1)} &=&
\frac{1}{m^2} \frac{1}{(4\pi)^2} \left(-\frac{\beta^3}{6}
(\varphi^\dagger\varphi)^3+ \frac{\beta^2}{12}
\partial_\mu (\varphi^\dagger\varphi)\partial^\mu (\varphi^\dagger\varphi)
\right.\nonumber\\
&&\left.
+\frac{g'^2\beta}{12} (\varphi^\dagger\varphi) B_{\mu \nu} B^{\mu \nu}-
\frac{g'^2}{60} \partial^\mu B_{\mu \nu} \partial_\sigma B^{\sigma\nu}
\right)\label{lagr1}~.
\end{eqnarray}
The coefficients of these operators  are free of divergences and
are suppressed by $1/m^2$.
Note that the last term in \eq{resdet},  which is trilinear in the field
strength, is zero in the Abelian case.

 It is clear that
 one could also obtain the interactions in \eq{lrenorm} and \eq{lagr1}
 by using ordinary Feynman rules. In fig.~2 we give the diagrams that
 give rise to this part of the effective  Lagrangian. Functional methods
 provide, however, a much more elegant and compact method to calculate
 them.

As mentioned before,  all the terms in  ${\cal L}_{ren}^{(1)}$
can be absorbed
in a redefinition of the standard model parameters.
The first term in \eq{lrenorm} is only a
renormalization of the vacuum energy. We can neglect it,
or, if we prefer, it can be absorbed in the constant term of the Higgs
potential of the standard model.  The second term renormalizes the
quadratic coupling in the Higgs potential:
\begin{equation}
\label{barmatchm}
\bar{m}^2_\varphi = m^2_\varphi - m^2
\frac{\beta}{(4\pi)^2}(1+\delte)~.
\end{equation}
The third one is a renormalization of
the quartic coupling of the standard Higgs:
\begin{equation}
\label{barmatchlambda}
\bar{\lambda} = \lambda - \frac{\beta^2}{2(4\pi)^2} \delte.
\end{equation}
Finally, the fourth term in \eq{lrenorm}, which is an additional
contribution to the  kinetic term for the
$B_\mu$-field, can be removed by  wave-function renormalization:
\begin{equation}
\label{barmatchb}
\bar{B}_\mu = \left(1+\frac{g'^2 \delte}{3(4\pi)^2}\right)^{1/2}
B_\mu~.
\end{equation}
In order to keep the canonical form of the covariant derivative,
we have to renormalize also
the $U(1)_Y$ coupling $g'$  keeping the product $g'B_\mu$
invariant:
\begin{equation}
\label{barmatchg}
\bar{g}'=  \left(1+\frac{g'^2 \delte}{3(4\pi)^2}\right)^{-1/2} g'~.
\end{equation}
The above equations \rfn{barmatchm}, \rfn{barmatchlambda}
 and  \rfn{barmatchg} relate the bare parameters of the full Lagrangian
$m_\varphi$, $\lambda$,  and $g'$
to the corresponding bare parameters
$\bar{m}_\varphi$, $\bar{\lambda}$ and $\bar{g}'$ of the effective Lagrangian.

\mysection{Matching the effective Lagrangian to the full Lagrangian}
\label{matching}

As commented in the previous section, it is necessary to expand the operator
of \eq{appsolution} in powers of $1/m^2$
 to have a local  effective
Lagrangian, but this expansion is only valid at the classical level,
since the limit of $m\rightarrow \infty$ and the functional integration
over the light fields do not commute. This is just a consequence of the fact
that the momentum $p$ of the scalar inside loop diagrams runs up to infinity
and then an expansion in $p^2/m^2$ is not appropriate. In spite of this
problem it is  still possible to obtain a local effective Lagrangian as
an  expansion in $1/m^2$ \cite{Wit76,Wit77}. It is necessary, however, to
include some additional operators that are not expected from the na\"{\i}ve
expansion in \eq{appsolution}. These
operators appear as a result of quantum corrections, hence, they are
suppressed by  additional couplings and $1/(4\pi)^2$ factors.

The practical way to obtain these  terms of the Lagrangian is
to consider the most general linear combination of all the operators
with  a given dimension, allowed by symmetry, and  then  to
extract the coefficients of this
combination by matching the effective Lagrangian calculation to the
full theory calculation.  Most of the operators obtained should
be included in any case as counterterms to cancel the  divergences
generated when the tree-level Lagrangian of \eq{lagr0} is used
inside loops. But it is important to note that
the finite parts of the coefficients of those operators can only
be fixed by computing the various Green functions with both the full and the
effective theories and matching the results for small energies compared
with the heavy mass.
Some other operators, however,  appear with finite coefficients. They are
just a consequence of the matching procedure and cannot be
obtained  from the divergences that appear in the effective theory.
In the diagrammatic language all these new operators are generated by
Feynman diagrams, with both the heavy scalar and lepton lines in the loops.
We will compute in the full theory the amplitudes corresponding to these
diagrams and will compare them with the equivalent amplitudes obtained
by using the tree-level effective Lagrangian, \eq{lagr0}, in one-loop
diagrams. The difference will give us the necessary counterterms.
 Actually, the effective Lagrangian  calculation can
be avoided by splitting the scalar propagator in two parts:
\begin{equation}
\label{splitting}
\frac{1}{k^2-m^2} = -\frac{1}{m^2}+\frac{1}{m^2} \frac{k^2}{(k^2-m^2)}\ .
\end{equation}
If we use this form when calculating diagrams with only one scalar propagator,
the first part gives
exactly what one would obtain by using the effective tree-level
four-fermion interaction, \eq{lagr0}, while the second part gives just
the counterterm we should add to the effective Lagrangian.
We would also like to note that by splitting the propagator in these
two pieces we have increased the degree of ultraviolet divergence in each of
the two terms, with respect to  the original diagram. On the other hand, the
 second
part contains an additional factor $k^2$ in the numerator, which reduces
the infared degree of divergence of this contribution. Therefore,
any possible small momentum singularity is transmitted to the
low-energy effective Lagrangian, as it should be, since infared
singularities have nothing to do with the high-energy behaviour of the
theory.

A  technical issue about the calculation is the selection of diagrams
one should compute to reconstruct the full effective Lagrangian. The
effective Lagrangian must be a linear combination of gauge-invariant
operators. Each of the operators gives rise to a variety of physical
processes. Obviously, to obtain the coefficients of these operators it is
not necessary to compute all these processes, since many
of them are related by gauge invariance. Our strategy consists in computing
all one-particle-irreducible diagrams with the minimal number of external
particles. We keep track
of all external momenta (of course only to the order $p^2/m^2$) without
using any equation of motion for the external particles. After that, we
write
an effective Lagrangian that reproduces those amplitudes and when
there is no ambiguity, we reconstruct gauge invariance by promoting
 any derivative to a covariant derivative.
Sometimes, however,
there is some ambiguity in the promotion of a derivative to a
covariant one. Then, we compute also the diagrams with one additional
external gauge boson in order to disentangle that ambiguity and we
check that the gauge-invariant effective Lagrangian correctly describes
all the amplitudes. Finally,  as an additional check,
we compute some diagrams with three external gauge bosons by
using both the full and the effective Lagrangians. We do it  for the
special case of
zero external momenta\footnote{We are studying a more efficient
method for the calculation of the gauge-invariant effective Lagrangian,
based on a calculation with constant fields similar to the one used in
appendix~\ref{det}.}.
Since the full calculation is tedious, we give the details
in  appendix~\ref{heavy-light}, presenting here only the main results
and an explanation of the procedure.

\subsection{Self-energies of the lepton-doublet}

We give in fig.~3 the diagrams contributing to the lepton wave function
renormalization in the full and the effective theories. In the effective
theory the wave function renormalization is zero because it is a massless
tadpole-like diagram (fig.~3.b).
However, in the full theory there is a non-zero
contribution that must be included as an additional effective operator.
{}From the results of  appendix \ref{heavy-light}
we find that, for the first term in \eq{selfe},
the additional operator we should include is
\begin{equation}
\label{selfenergy}
 2(\delte+\frac{1}{2}) i (\overline{\ell} F \db \ell)\ ,
 \end{equation}
 where we have defined the matrix $F$ as
\begin{equation}
\label{defF}
F_{ab} \equiv \frac{ (f^\dagger f)_{ab}}{(4\pi)^2}\ .
\end{equation}
 Note that even though in  appendix~\ref{heavy-light}
 we have only calculated the charged scalar contribution to the
self-energy of the doublet,
gauge invariance requires that the partial derivative should
be substituted by a covariant derivative. In this case the promotion
from the derivative to the covariant derivative can be done without
ambiguity. This leads to \eq{selfenergy}. It  implies that there
should be an additional contribution to the coupling of the gauge bosons to
the lepton doublet given by the coefficient in \eq{selfenergy}.
We will see in the next subsection that, indeed, this contribution appears.

The second term in \eq{selfe} is proportional to $\sla{p} p^2$, which
requires an effective Lagrangian of the form
$ i (\overline{\ell}_a \sla{\partial} \partial^2 \ell_b)$.
However, the promotion of this term to covariant derivatives is
ambiguous: should we use $\db D^2$, $\db^3$ or $D_\mu \db D^\mu$ ? The
only way to resolve this ambiguity, as we will see below,
is to perform a full calculation with one external gauge boson.

The operator in
 \eq{selfenergy}
contributes to the kinetic term of the doublets. To write it again in
a canonical form, we should redefine the lepton doublet as follows:
\begin{equation}
\label{wfren}
\ell_a \rightarrow
\ell_a -\frac{(\delte+\frac{1}{2})}{(4\pi)^2}
(f^\dagger f)_{ac} \ell_c\ .
\end{equation}
This redefinition only affects the tree-level four-fermion
effective Lagrangian and the standard model Yukawa coupling. This is
because
the wave function renormalization
appears at the one-loop level, and, as a consequence, its effect on all
the other
(one-loop) operators is a two-loop effect. It
can be taken into account by redefining the standard model Yukawa couplings
and the couplings $f_{ab}$ that appear in \eq{lagr0}  as follows
\begin{eqnarray}
Y_{ab} \rightarrow & \bar{Y}_{ab} = & Y_{ab} -
\frac{(\delte+\frac{1}{2})}{(4\pi)^2} (f^\dagger f Y)_{ab}\label{redefY}\ ,\\
f_{ab} \rightarrow & \bar{f}_{ab} = & f_{ab} -
2\frac{(\delte+\frac{1}{2})}{(4\pi)^2}
(f f^\dagger f)_{ab} \label{redeff}\ .
\end{eqnarray}
As we see this does not change the flavour structure of the
couplings, since the $\bar{f}_{ab}$ is still antisymmetric in flavour
indices and
does not give any new interesting process.  The same thing could be
said about the standard model Yukawa couplings $\bar{Y}_{ab}$.

\subsection{Penguins}
In fig.~4 we present the contributing diagrams to the vertex of the
 $B_\mu$ and $\vec{W}_\mu$ gauge bosons with leptons, with the heavy scalar
running in the loop (figs.~4.a and 4.b). Note that for the
$\vec{W}_\mu$ there is no diagram fig.~4.b,
because  the scalar is an $SU(2)$ singlet.
We also give the corresponding
diagrams in the effective theory (fig.~4.c). Here it is interesting to note
that the amplitude corresponding to the diagrams of figs.~4.a  and
4.b in the full theory
is finite,  when the diagrams
with external wave function renormalization are included.
This can be understood because the full theory is renormalizable:
since all these diagrams lead to a flavour-changing vertex and
as our  original Lagrangian does not contain such a coupling,
there is no  counterterm available.
In the effective theory language,
this cancellation is just a consequence of gauge
invariance: the divergent contributions to the vertex in the full
theory can be taken into account by the gauge-invariant dimension-four
operator of \eq{selfenergy}, which can be
removed by a wave function renormalization of the lepton doublet.

We give in  appendix~\ref{heavy-light} the resulting calculation,
eqs.~\rfn{ampB} and \rfn{ampW},
of  the vertex diagrams of the $B_\mu$ and $\vec{W}_\mu$ after subtraction of
the effective Lagrangian contribution (diagram 4.c). We cast this
result in a form that can be easily identified with some effective
operators. The first term can  already be obtained, as promised,
from \eq{selfenergy}. The rest of the terms can be obtained (for both
the $B_\mu$ and the $\vec{W}_\mu$ fields) from the
following operators:
\begin{eqnarray}
&&-\frac{1}{3m^2}   i (\overline{\ell}  F (D^2 \db+ \db D^2 )\ell)
\label{selfenergy2}\\
&&+ \frac{2}{3m^2} \left(\delte+\frac{5}{3}\right)
\frac{g'}{m^2}  (\overline{\ell} F \gamma_\nu \ell) D_\mu B^{\mu \nu}
+ \frac{2}{3m^2}\left(\delte+\frac{4}{3}\right)
 \frac{g}{m^2}  (\overline{\ell} F \gamma_\nu \vec{\tau} \ell)
 D_\mu \vec{W}^{\mu \nu}\hspace{1.5cm}\label{penguin1}\\
&&+\frac{g'}{4m^2} B^{\mu \nu}
(\overline{\ell} F i \sigma_{\mu\nu} \db \ell+ \mrm{h.c.})
+\frac{g}{4m^2} \vec{W}^{\mu \nu}
 (\overline{\ell} F i \sigma_{\mu\nu} \vec{\tau} \db\ell+\mrm{h.c.})
 \label{penguin2}\ .
\end{eqnarray}
In these equations,  $D_\mu$ is the appropriate covariant derivative for each
of the fields. It is easy to see that the pure derivative part of the first
term
\eq{selfenergy2},  reproduces perfectly
the second term in \eq{selfe} and that the ambiguity  mentioned in
the previous subsection has been solved completely with this calculation.
To check  this result further,  we have computed the diagram
with three $W_\mu$ gauge bosons for zero external momenta and
compared the result with that obtained from
eqs.~(\ref{selfenergy2})--(\ref{penguin2}) for constant fields.

The physical content of these operators is quite obscure.
Then, we found it convenient to use the operator identity:
\begin{eqnarray}
\label{relation}
\db^3 &=& \frac{1}{2} (D^2\db+\db D^2)\\ \nonumber
&&+
\frac{1}{8} \left(
g' \sigma_{\mu\nu} B^{\mu\nu}\db+g' \db \sigma_{\mu\nu}  B^{\mu\nu}-
g \sigma_{\mu\nu} \vec{W}^{\mu\nu} \vec{\tau} \db
-g \db \sigma_{\mu\nu}  \vec{W}^{\mu\nu} \vec{\tau} \right)
\end{eqnarray}
to rewrite eqs.~(\ref{selfenergy2})--(\ref{penguin2})
in the following form:
\begin{eqnarray}
&&-\frac{2}{3m^2}   i (\overline{\ell}  F  \db^3\ell)
\label{selfenergy2a}\\
&&+ \frac{2}{3} \left(\delte+\frac{5}{3}\right)
\frac{g'}{m^2}   (\overline{\ell} F \gamma_\nu \ell) D_\mu B^{\mu \nu}
+ \frac{2}{3}\left(\delte+\frac{4}{3}\right)
 \frac{g}{m^2}  (\overline{\ell} F \gamma_\nu \vec{\tau} \ell)
 D_\mu \vec{W}^{\mu \nu}
 \label{penguin1a}\\
&&+\frac{g'}{3m^2} B^{\mu \nu}
(\overline{\ell} F i \sigma_{\mu\nu} \db \ell+ \mrm{h.c.})
+\frac{g}{6m^2} \vec{W}^{\mu \nu}
 (\overline{\ell} F i\sigma_{\mu\nu} \vec{\tau} \db\ell+\mrm{h.c.})\ .
 \hspace{1cm}\label{penguin2a}
\end{eqnarray}

As we will see later, after using the equations of motion,
the physical content of
these operators
becomes more transparent. The first term  can be removed
in favour of Yukawa-type couplings and does not give any interesting
process. The second and the third terms will give rise to
processes such as $Z\rightarrow  \mu e$ and $\mu \rightarrow e e e$,
and finally the last terms, with the appropriate combination of
$W^3_\mu$ and $B_\mu$, will give rise to $\mu \rightarrow e \gamma$.

\subsection{Seagull diagrams}

There are also some operators that involve two scalar Higgses,
two lepton doublets, and one covariant derivative. They are generated
by the diagrams of  fig.~5 and they do not contribute to
any interesting process. For completeness we give  the result here.
{}From the diagram of fig.~5.b we get the amplitude \eq{amp5a},
 which can be obtained by using  the  operator
\begin{equation}
\label{op5a}
i \frac{\beta}{m^2} (\varphi^\dagger \varphi)
(\overline{\ell} F \db \ell) + \mrm{h.c.}\ .
\end{equation}
{}From the diagram of fig.~5.a, and after subtraction of the effective
Lagrangian
contribution, fig.~5.c, we get the amplitude \rfn{amp5b}, which can be obtained
from the operators
\begin{eqnarray}
&&(\delte+1)\frac{i}{m^2} \left((D_\mu \varphi)^\dagger \varphi\right)
(\overline{\ell} \hat{F} \gamma^\mu \ell)
+(\delte+1) \frac{i}{m^2} \left((D_\mu \varphi)^\dagger
\vec{\tau}\varphi\right)
(\overline{\ell} \hat{F} \gamma^\mu \vec{\tau} \ell)\nonumber
\hspace{2cm} \\
&&-\frac{i}{2m^2} \left(\varphi^\dagger \varphi\right)
(\overline{\ell} \hat{F} \db\ell_b)
-\frac{i}{2m^2} \left(\varphi^\dagger \vec{\tau}\varphi\right)
(\overline{\ell} \hat{F} \vec{\tau} \db \ell)+\mrm{h.c.}\label{op5b}\ ,
\end{eqnarray}
where
\begin{equation}
\label{defhatF}
\hat{F}_{ab} = \frac{(f Y_e Y^\dagger_e f^\dagger)_{ba}}{(4\pi)^2}\ .
\end{equation}
We used an $SU(2)$ Fierz transformation to write the operators in
this form, being the original operators expressed in terms of
$\tilde{\varphi}$. Clearly the effects of operators \rfn{op5b}
are suppressed
by at least two powers of the lepton masses.

\subsection{Four-fermion interactions}

In this section we shall discuss possible new four-fermion interactions
that could arise at the one-loop level. These interactions
can come from three different kinds of diagrams:

i) Diagrams with gauge boson corrections to the vertex of the charged
scalar with the lepton doublet.  Since these only contain light particles
in the loop, they are fully taken into account by gauge boson corrections
to the effective four-lepton interaction \rfn{lagr0}.  In addition they do not
change the flavour structure of the coupling.

ii) Box diagrams with   a charged scalar and a gauge boson running in
the loop (fig.~6.a).

It is easy to see that diagrams with a $\vec{W}_\mu$
exchange just cancel because of the antisymmetry of the couplings $f_{ab}$.
Diagrams with exchange of $B_\mu$
renormalize the tree-level four-lepton interaction without changing
its flavour structure.  Part of their contributions are taken into account
by the effective theory diagrams of  fig.~6.c. But, obviously,
these diagrams
do not modify the flavour structure of the tree-level effective
four-fermion interaction (\ref{lagr0}) and, then, they
cannot give rise to processes such as  $\mu \rightarrow e e  e$
at least at order $1/m^2$.

iii) Box diagrams with two charged scalars in the loop
(fig.~6.b).  Diagrams of this type are finite and in the sum they give
rise to the following operator:
\begin{equation}
\label{boxb}
-\frac{(4\pi)^2}{m^2} F_{ab} F_{cd} (\overline{\ell}_a \gamma_\mu \ell_b)
(\overline{\ell}_c \gamma^\mu \ell_d)\ .
\end{equation}
It can be shown that this operator does not contribute to processes
with two identical fermions, such as $\mu^- \rightarrow e^+ e^- e^-$.
Indeed,  taking only into account, for example, the down component
of the $SU(2)$ lepton doublets, the interaction among four charged
leptons induced by \eq{boxb} can be written as
\begin{equation}
 \label{boxbb}
-\frac{(4\pi)^2}{m^2} F_{ab} F_{cd} (\overline{e_{aL}} \gamma_\mu e_{bL})
(\overline{e_{cL}} \gamma^\mu e_{dL}) =
-\frac{(4\pi)^2}{2m^2} ( F_{ab} F_{cd}- F_{ad} F_{cb})
 (\overline{e_{aL}} \gamma_\mu e_{bL})
(\overline{e_{cL}} \gamma^\mu e_{dL})\ ,
  \end{equation}
where, to obtain the right-hand side, we have used a Fierz transformation.
We immediately see that this operator does not contribute
to processes
such as  $\mu^- \rightarrow e^+ e^- e^-$. However it does contribute to other
interesting processes such as $\tau^- \rightarrow e^+ \mu^- e^-$.

Clearly only the contributions of  type (iii) are interesting and only
these will be included in our effective Lagrangian.

 \vspace{0.5cm}
 Before we summarize the results of section~4, we
note that all the results obtained in section~\ref{matching}
have been derived from the effective tree-level Lagrangian
given in \eq{lagr0}. One can easily see that this Lagrangian can also be
written
(after a combined Fierz transformation in $SU(2)$ and Dirac space) in the
more familiar form
\begin{equation}
{\cal L}^{(0)}\  ``="\  \frac{1}{m^2} f_{ab} f^*_{cd}
(\overline{\ell}_d \gamma^\mu \ell_b)
(\overline{\ell}_c \gamma_\mu \ell_a)\ .
\end{equation}
But this is true only  in four dimensions: the Fierz transformations we used
above cannot be
done in $D$ dimensions. Should we start with this new Lagrangian,
the finite parts of the counterterms would be different when using
dimensional regularization, but this formulation would be completely
equivalent to the original one. We decided to keep the original
form of the Lagrangian so as to keep the notation as close as possible to the
full Lagrangian formulation.

 Using the results of sections 3 and 4, the complete
 one-loop effective Lagrangian is
\begin{equation}
{\cal L}_{eff} = {\cal L}_{SM} +  {\cal L}^{(0)}+ {\cal L}_{det}^{(1)}
+{\cal L}_{match}^{(1)}\ .
\end{equation}
 Here ${\cal L}_{SM}$ is the standard model Lagrangian, \eq{smlagrangian},
 ${\cal L}^{(0)}$ as given in \eq{lagr0} is the tree-level $1/m^2$ term
 of the effective Lagrangian, ${\cal L}_{det}^{(1)}$ is given in \eq{lagr1}
 and contains all $1/m^2$ one-loop contributions of diagrams with only
 singlet scalars in the loop and finally ${\cal L}_{match}^{(1)}$, which is
 the sum of the terms in eqs.~(\ref{selfenergy2a})--(\ref{op5b}) and
 \rfn{boxb},
 gives all the
 contributions coming
 from matching the full theory in diagrams with heavy-light particles
 in the loops:
\begin{eqnarray}
{\cal L}_{match}^{(1)}
&=& \frac{2}{3} (\delte+\frac{5}{3})
\frac{g'}{m^2} (\overline{\ell} F\gamma_\nu \ell) D_\mu B^{\mu \nu}
+ \frac{2}{3}(\delte+\frac{4}{3})
 \frac{g}{m^2}
 (\overline{\ell} F\gamma_\nu \vec{\tau} \ell) D_\mu \vec{W}^{\mu \nu}
 \nonumber \\
&&
+\ \frac{1}{3} \frac{g'}{m^2} B^{\mu \nu}
(\overline{\ell} F i \sigma_{\mu\nu} \db \ell+ \mrm{h.c.})
+\frac{1}{6} \frac{g}{m^2} \vec{W}^{\mu \nu}
 (\overline{\ell} F i\sigma_{\mu\nu} \vec{\tau}
 \db\ell+\mrm{h.c.})\hspace{1cm}\nonumber \\
 && +\left(i\frac{\beta}{m^2} (\varphi^\dagger \varphi)
(\overline{\ell} F\db \ell) + \mrm{h.c.} \right)
-\frac{2}{3m^2}   i (\overline{\ell} F \db^3\ell)
\nonumber\\
&& +\left(\left.\right. (\delte+1) \frac{i}{m^2}
\left((D_\mu \varphi)^\dagger \varphi\right)
(\overline{\ell} \hat{F}\gamma^\mu \ell)
+(\delte+1) \frac{i}{m^2} \left((D_\mu \varphi)^\dagger
\vec{\tau}\varphi\right)
(\overline{\ell} \hat{F}\gamma^\mu \vec{\tau} \ell)\right.\nonumber\\
&& \left.
-\frac{i}{2m^2} \left(\varphi^\dagger \varphi\right)
 (\overline{\ell}\hat{F}\db\ell)
-\frac{i}{2m^2} \left(\varphi^\dagger \vec{\tau}\varphi\right)
(\overline{\ell} \hat{F}\vec{\tau} \db \ell)+\mrm{h.c.} \right)\nonumber\\
&&
-\frac{(4\pi)^2}{m^2}  (\overline{\ell} F \gamma_\mu \ell)
(\overline{\ell} F \gamma^\mu \ell) \ . \label{totalefflagr}
\end{eqnarray}
 All the couplings in this Lagrangian should be understood as
 bare barred effective Lagrangian couplings related to the full theory
 bare couplings  through eqs.~\rfn{barmatchm}, \rfn{barmatchlambda},
 \rfn{barmatchg},  \rfn{redefY} and \rfn{redeff}.
To simplify the notation we have
suppressed everywhere generational indices, then
 $F$ and $\hat{F}$ are the matrices
defined in eqs.~(\ref{defF}) and (\ref{defhatF}) .

\mysection{Renormalization and operator mixing analysis\newline
 of the effective
Lagrangian}
 \label{renormalization}

Equations~\rfn{barmatchm}, \rfn{barmatchlambda}, \rfn{barmatchg},
 \rfn{redefY} and \rfn{redeff} relate the bare
couplings and masses of the
effective theory with those
of the full theory. It is, however, more interesting to have the
equivalent relations for renormalized couplings.
If an $\overline{MS}$ scheme\footnote{It has been customary in the
recent literature to use cut-off regularization schemes when working
with effective Lagrangians and then  identify the cut-off with the
scale of new physics.
The interpretation of the ultraviolet
cut-off as the scale of new physics has led to some erroneous results.
The effective Lagrangian should also be renormalized, and
physical results should be independent of the regularization and of
the renormalization schemes.
We will see later that
dimensional regularization with minimal subtraction leads to the same
physical results   in a cleaner way;
therefore,  we will always use this scheme.} is
employed to renormalize both the full and the effective theories,
we can very easily obtain  the matching equations for the
renormalized couplings. Let us take, for example, the gauge coupling $g'$.
We
denote the $\overline{MS}$ renormalized quantities with the same symbol as
the bare quantities, but adding an additional dependence on the
renormalization scale $\mu$.  All effective theory quantities will be
distinguished by a bar.  The standard relationship between bare and
renormalized couplings in $\overline{MS}$ schemes is  ($D=4-2\epsilon$
and $\frac{1}{\hat{\epsilon}}= \frac{1}{\epsilon}-\gamma+\log (4\pi)$):
\begin{eqnarray}
g' \mu^\epsilon &=& g'(\mu)+ \frac{1}{2\hat{\epsilon}}\ b_{g'}
g'^3(\mu)+\cdots\ , \nonumber\\
\bar{g}' \mu^\epsilon &=& \bar{g}'(\mu)+
\frac{1}{2\hat{\epsilon}}\ \bar{b}_{g'} \bar{g}'^3(\mu)+\cdots\ ,\nonumber
\end{eqnarray}
where $b_{g'}$ and $\bar{b}_{g'}$ are the lowest-order coefficients of the
$\beta$-functions for the coupling constants in the full and  the effective
theories, respectively.  Substituting these equations in
\eq{barmatchg} and equating finite terms \cite{Wei80,Hal81} we obtain
the desired matching condition for renormalized couplings:
\begin{equation}
\label{matchg}
\bar{g}'(\mu) = g'(\mu) - \frac{g'(\mu)^3}{3(4\pi)^2} \log(\mu/m)+\cdots\
\  .
\end{equation}
Note that this equation can be obtained by just  dropping the
$1/\hat{\epsilon}\ $ contained in $\delte$ in
\eq{barmatchg}. This is not surprising since the divergent term in
\eq{barmatchg} gives just the charged scalar contribution to the
beta function of $g'$ in the full theory.
Similar arguments give for the rest of
 the
couplings and masses the following result
\begin{equation}
\label{matchm}
\bar{m}^2_\varphi(\mu) = m^2_\varphi(\mu) - m^2(\mu)
\frac{\beta(\mu)}{(4\pi)^2}(1+2\log(\mu/m))\ ,
\end{equation}
\begin{equation}
\label{matchlambda}
\bar{\lambda}(\mu) = \lambda(\mu) -
\frac{\beta^2(\mu)}{(4\pi)^2}\log(\mu/m)\ ,
\end{equation}
\begin{equation}
\label{matchY}
\bar{Y}_{ab}(\mu) =  Y_{ab}(\mu) -
\frac{(f^\dagger f Y)_{ab}(\mu)}{(4\pi)^2}
(\frac{1}{2}+2\log(\mu/m))
\end{equation}
and a matching equation for the four-lepton coupling in the effective
theory expressed in terms of the full theory couplings and masses.
It can be obtained by using an
effective $\bar{f}_{ab}(\mu)$ defined as follows:
\begin{equation}
\label{matchf}
\bar{f}_{ab}(\mu) = f_{ab}(\mu) -
2 \frac{(f f^\dagger f)_{ab}(\mu)}{(4\pi)^2}
\left(\frac{1}{2}+2\log(\mu/m)\right)
+\cdots\ \ .
\end{equation}
In this equation there could also be some contributions from box diagrams,
which we have not computed.

These matching conditions are, in principle, valid for an arbitrary value of
the
renormalization scale $\mu$.   However, it is clear that
in order to avoid large logarithms they  should be evaluated at some scale
around the charged scalar mass and then, using the standard model
renormalization group, run all the couplings to obtain their values at
lower scales.

Equation \rfn{matchm} is very interesting, since it manifests in all its
crudeness the so-called naturalness problem of the standard model;
$\bar{m}_\varphi(\mu)$ is the mass parameter that appears in the
Higgs potential part of the effective Lagrangian, and it has to be of
the order of the
electroweak scale. However, it is clear that if we take $m(\mu)$ very
large, we should also take $m_\varphi(\mu)$ large in order to have
$\bar{m}_\varphi(\mu)$ small enough. But even if we do so at some
scale $\mu$, it will be very difficult to keep $\bar{m}_\varphi(\mu)$
small at any other scale. This represents a serious fine-tuning
problem, which appears when the standard model is embedded in
another model containing mass  scales much larger than the Fermi scale.
It is important to note that by using dimensional regularization the
problem appears only in the matching conditions. If a cut-off
regularization scheme is used, the naturalness problem can be related to
the appearance of quadratic divergences.
However, this relation is not direct. Quadratic
divergences can appear at the intermediate stages, but
they should be absorbed in a renormalization of the mass
parameters of both the full and the effective
theories, since physical quantities must be cut-off-independent.
After renormalization, the naturalness problem should
also appear only as a fine-tuning problem in the matching conditions,
as in \eq{matchm}.  As the main physical consequences can be obtained
equally in dimensional regularization, we do not find any particular
advantage by working with a  cut-off regularization scheme.

If an $\overline{MS}$
scheme is used to renormalize also the higher dimension operators,
we could write the effective Lagrangian in
terms of a set of running couplings $c_i(\mu)$ for each of the operators
\begin{equation}
{\cal L}_{eff} = \sum_i  c_i(\mu) O_i\ ,
\end{equation}
where the $c_i(\mu)$ are obtained from the bare coefficients
in \eq{totalefflagr} just
by dropping the $1/\epsilon -\gamma+\log(4\pi)$ terms in
$\delte$; that is just by substituting $\delte$ by $2\log(\mu/m)$.
Then all
couplings should be understood as renormalized effective Lagrangian
couplings (barred couplings) related to the full theory couplings through
eqs.~(\ref{matchg})--(\ref{matchf}).

To  simplify matters we can consider only processes that violate
muon-lepton number in one unit and electron-lepton number in minus one
unit (the total lepton number must be conserved).  For example, we can
take the two operators
\begin{eqnarray}
O_1  &\equiv &(\elltbar_3  \ell_2)
 (\overline{\ell}_1  \widetilde{\ell}_3) \nonumber\\
O_2  & \equiv & \frac{1}{g'} D_\mu B^{\mu \nu}
(\overline{\ell}_1 \gamma_\nu \ell_2)  \nonumber
\end{eqnarray}
The factor $1/g'$ in the penguin operator has
been included for convenience.
 From eqs.~\rfn{lagr0} and \rfn{penguin1}, we obtain expressions
 for the corresponding running couplings:
 \begin{eqnarray}
 c_1(\mu) &=& \frac{(f^\dagger f)_{12}(\mu)}{m^2} + \cdots \label{cesp}\\
 c_2(\mu) &= &
 \frac{(f^\dagger f)_{12}(\mu)}{m^2} \frac{4 g'^2(\mu)}{3(4\pi)^2}
  \log\frac{\mu}{m}
+ \frac{(f^\dagger f)_{12}(\mu)}{m^2} \frac{10 g'^2(\mu)}{9(4\pi)^2}
+\cdots\ , \label{ces}
  \end{eqnarray}
 where the dots represent additional one-loop contributions, which, as
 we will see immediately, are not important at the level we are working.
 Note that $(f^\dagger f)_{12} = f^*_{31} f_{32}$ because of the
 antisymmetry of the couplings in flavour indices.
 Equations~(\ref{cesp})--\rfn{ces} can be cast into the following form
\begin{eqnarray}
c_1(\mu) &=& c_1(m)\left(1+\gamma_{11} \log\frac{\mu}{m}\right)
+c_2(m) \gamma_{12} \log\frac{\mu}{m}  \label{c1}\\
c_2(\mu) &=&  c_1(m) \gamma_{21}   \log\frac{\mu}{m}
+ c_2(m) \left(1+\gamma_{22}\log\frac{\mu}{m}\right) \ ,\label{c2}
\end{eqnarray}
 with
\begin{equation}
\label{gammas}
\gamma_{11} \approx  O\left(\frac{g^2}{(4\pi)^2}\right) ,\ \ \
\gamma_{12} \approx O\left(\frac{g^2}{(4\pi)^2}\right),\ \ \
\gamma_{21}
=\frac{4}{3}\frac{g'^2 }{(4\pi)^2},\ \ \
 \gamma_{22} \approx O\left(\frac{g^2}{(4\pi)^2}\right)
 \end{equation}
and the couplings at the scale $\mu=m$ are given by
\begin{equation}
\label{boundary}
c_1(m) = \frac{(f^\dagger f)_{12}(m)}{m^2} ,\bla c_2(m) =
\frac{10}{9}\frac{g'^2}{(4\pi)^2} \frac{(f^\dagger f)_{12}(m)}{m^2} \ .
\end{equation}
 Since $c_2(m)$  appears only at the one-loop level
 in the full model, the last term in \eq{c1} is a two-loop effect.
 This is the reason why we
 have not considered it in \eq{cesp}. On the other hand, if $\mu$ is not
 very different from $m$ we have $\gamma_{11} \log(\mu/m)  \ll 1$ and
 $\gamma_{22} \log(\mu/m)  \ll 1$; then,  the diagonal elements
 of the anomalous dimension matrix do not need to be computed.

 Equations~\rfn{c1} and \rfn{c2} are an approximate solution of the general
 renormalization group equation that describes mixing among
 the operators $O_1$ and $O_2$
\begin{equation}
\label{renopmix}
\mu \frac{ d c_i(\mu)}{d\mu} = \gamma_{ij} c_j(\mu)\ ,
\end{equation}
which is valid only in the case that $\gamma_{ij} \log(\mu/m) \ll 1$. If
we pretend to use this solution for a wider range of $\mu$'s,
 the full $\gamma_{ij}$
matrix should  be computed and the integration of \eq{renopmix}
should be done by taking also into account the running of the gauge
coupling constants. However, for the standard model values of the
gauge coupling constants the linear approximation
($\gamma_{ij}~\log(\mu/m)\ll~1$) works very well
for a  wide range of scales
$(\gamma_{ij} \sim 10^{-3})$. Obviously, in this approximation and
taking into account that only the operator in \eq{lagr0} is generated
at tree level it is clear that in our model it is enough to consider
only mixing of the effective operators generated at the one-loop level with
the tree-level operator because other mixings would represent two-loop
effects.
Then, at the order we are working, we have always $2\times 2$ operator
mixing.

This simple example shows us clearly what can and what cannot
be obtained by using
the effective Lagrangian.
{}From the effective theory we could calculate the anomalous
dimension matrix that controls the mixing among the different
operators in the effective Lagrangian since the logarithmic
terms are the same in the full and in the effective theories,
however the boundary conditions for the
renormalization group equation (\ref{renopmix}) can only
be obtained after the matching procedure. If we do not
know the full theory a set of boundary conditions can only be obtained
from experiment.  In this case we can still compute
the anomalous dimensions in the effective theory, write
down  \eq{renopmix}, and solve it for arbitrary
boundary conditions at the scale $\mu_0$. Then the solution, within
our previous approximations, is
\begin{eqnarray}
\label{ces3}
c_1(\mu) &=&
c_1(\mu_0)\left(1+\gamma_{11} \log\frac{\mu}{\mu_0}\right)
+c_2(\mu_0) \gamma_{12} \log\frac{\mu}{\mu_0}\nonumber\\
c_2(\mu) &=&  c_1(\mu_0) \gamma_{21}   \log\frac{\mu}{\mu_0}
+ c_2(\mu_0) \left(1+\gamma_{22}\log\frac{\mu}{\mu_0}\right)\ .
\end{eqnarray}
Now we should take the full, complete basis of operators that
mix at the one-loop level, since if the boundary conditions are not known
we cannot neglect {\it a priori} any of the mixings. To keep simplicity,
however,
we will discuss only two-operator mixing.
The interesting point about eqs.~\rfn{ces3}
is that, if we know all the effective
couplings at some scale $\mu_0$, we can obtain them at some other scale
(not necessarily the scale responsible for the new physics).
Equations~\rfn{ces3} clearly tell us that to predict the values of the
couplings at any scale we need two (in the case of mixing of two operators)
boundary conditions. These conditions could be obtained from experiment.
For example, if we could bound the couplings of the two operators at LEP1
we could straightforwardly predict these couplings at LEP2 energies.
However,  if in our example we only know one
of the couplings at LEP1,  no matter how many loops we use to calculate
the anomalous dimensions, we will not
be able to bound the other operator at the same or any other scale.
Only if the full theory is known  can we relate the couplings of different
operators as they are expressed in terms of the few parameters of the full
theory. Then, bounds on
one operator can be related to bounds on other operators.
In our example it is clear from \eq{boundary} that
\begin{equation}
\label{crelation1}
 c_2(m) = \frac{10}{9}\frac{g'^2}{(4\pi)^2} c_1(m)
\end{equation}
or,  for instance at the $M_Z$ scale,  by using eqs.~\rfn{cesp}--\rfn{ces}
we have
\begin{equation}
\label{crelation2}
 c_2(M_Z) =
 \left(\frac{4 g'^2(M_Z)}{3(4\pi)^2} \log(M_Z/m)
 +\frac{10 g'^2(M_Z)}{9(4\pi)^2}\right) c_1(M_Z)\ .
 \end{equation}
 Then it is clear that a bound on $c_1(M_Z)$ implies a bound on
 $c_2(M_Z)$, and viceversa,  and it also implies a bound on $c_1(\mu)$ and
 $c_2(\mu)$ at any other scale $\mu$. But it is important to realize
 that this  is only possible because our full theory provides us with
 additional relationships among couplings.
  Of course, one can always impose additional assumptions to obtain
 an estimate of the bounds on different operators.

 For example,
 if one assumes that there
 are no unnatural cancellations among the couplings of different operators
 one can bound each of them independently from the others.  In our example
 this would be implemented by putting, for instance, $c_1(m) \not= 0$ and
 $c_2(m) = 0$, then use eqs.~\rfn{c1} and \rfn{c2} with $\mu=M_Z$
 to obtain $c_1(M_Z)$ and $c_2(M_Z)$ as a function of $c_1(m)$.
 Clearly with this assumptions a bound on $c_2(M_Z)$ implies a bound
 on $c_1(m)$ and therefore also a bound on $c_1(M_Z)$. And similarly for
 the other coupling.
 The estimates so obtained are interesting and
 can be useful. However,  we feel that they
 can never substitute more elaborate bounds  based on
 a complete set of experimental data.

\mysection{Spontaneous symmetry breaking and\newline
 the use of the equations
of motion}
\label{ssb}

{}From all  previous sections  it is clear that our effective Lagrangian
reproduces the results of the full theory at the one-loop level
as long as energies smaller than the $h$ mass are involved.  However,
until now we have considered the full unbroken theory, but the standard
model and the extension of it we are considering now undergo spontaneous
symmetry breaking (SSB) and this has to be taken into account in the analysis.

 Another point that has been the origin of controversy is the possibility of
using the standard model equations of motion  to reduce the number
of operators in the effective Lagrangian or to make its
physical content more evident. It is well known that the
equations of motion cannot be used na\"{\i}vely in a Lagrangian if it
is going to be used to calculate diagrams with particles off-the-mass-shell.
In fact, the possibility of using the equations of motion to trade
derivative couplings in favour of pseudo-scalar couplings
has led to many wrong calculations in pion physics
and in other models containing Goldstone bosons.  However,
the equations of motion
can still be used under certain circumstances. First of all, if the
Lagrangian is going to be used to calculate tree-level amplitudes with
all particles on-the-mass-shell, they can be used without any problem.
Moreover
it can be shown \cite{Geo91,Pol80,Arz92} that the use of the equations of
motion is equivalent
to a redefinition (in general with a non-linear transformation)
of the fields of the theory. If the starting theory is  renormalizable,
normally  these non-linear
transformations lead to Lagrangians that contain operators of
higher dimension
and which are not explicitly renormalizable.
However,  if the starting
theory is described by  an effective
Lagrangian,  with all kinds of non-renormalizable
interactions, the effect of these transformations is equivalent
to the use of the equations of motion
plus a modification of the higher-order terms
in the effective Lagrangian. This  is not a problem since
all these terms are already present in the effective Lagrangian. Then,
it only amounts to a re-ordering
of the effective Lagrangian \cite{Geo91}. There are some subtleties related to
the Jacobian of the transformation and renormalization, but in general
it can be shown that they do not represent a problem \cite{Pol80,Arz92}.
However,  one has to be careful when studying processes that are
sensitive to operators of different dimensions.  In our case there is
no problem since the only  operators that could be eliminated with
the equations of motion appear at the one-loop level;  then,  they are
going to be used uniquely as tree-level insertions.
We could use the equations of motion
before or after SSB, or  not use them at all, but the
final result should be independent from this.  Intermediate steps, however, can
look quite different.
We found it convenient to use the equations of motion of the standard
model before SSB, because
they are simpler and because
it makes it easier to see which
new physics is generated by the effective Lagrangian.
 We are going to use them
 to substitute the derivatives of field strengths
 in favour of fermionic currents and new gauge boson interactions
 involving the standard Higgs scalar. Similarly, we will substitute
 covariant derivatives of the lepton doublet in favour of the right-handed
 leptons and the standard Higgs.  Later on we will
see, in a particular example, that physical observables are the same as
the results one would have obtained if proceeding in a different way.

{}From the standard model Lagrangian, \eq{smlagrangian}, one can easily
obtain the following equations of motion for the electroweak gauge bosons
\begin{eqnarray}
D_\mu B^{\mu \nu}& = & - g' \left(J^\nu_B+
\frac{i}{2}  \varphi^\dagger \lrover{D}^\nu \varphi\right) \label{eqmotB}\\
D_\mu \vec{W}^{\mu \nu} & = &- g \left(\vec{J}^\nu_W+
 \frac{i}{2} \varphi^\dagger \lrover{D}^\nu \vec{\tau}\varphi\right)
 \label{eqmotW}\ ,
 \end{eqnarray}
where
\begin{eqnarray}
 J^\nu_B &=&
-\frac{1}{2} \bar{\ell} \gamma^\nu \ell
-\bar{e} \gamma^\nu e
+\frac{1}{6} \bar{q} \gamma^\nu q
+\frac{2}{3} \bar{u} \gamma^\nu u
-\frac{1}{3} \bar{d} \gamma^\nu d \label{currentB}\\
\vec{J}^\nu_W &=&
\frac{1}{2}\bar{\ell} \gamma^\nu \vec{\tau} \ell
+\frac{1}{2}\bar{q} \gamma^\nu \vec{\tau} q \label{currentW}
\end{eqnarray}
are the hypercharge and $SU(2)$ fermionic currents\footnote{Note that
eqs.~\rfn{eqmotB}--\rfn{currentW} differ from eqs.~(2.13) and (2.14)
in
ref.~\cite{BW86}, where some terms are missing.}.
We are also going to use the standard model equation of motion for
the leptonic doublet
\begin{equation}
\label{eqmotdoublet}
i \db \ell=- Y_e e \varphi\ .
\end{equation}

We will use eqs.~\rfn{eqmotB}--\rfn{eqmotW} and \rfn{eqmotdoublet} in
(\ref{totalefflagr}). Every time we use the
equation of motion for the leptonic doublet we obtain an additional
leptonic Yukawa coupling constant $Y_e$. Since  these couplings are
small, we are going to neglect terms that contain two or more of those
Yukawa couplings.

After using  \rfn{eqmotB}, \rfn{eqmotW} and \rfn{eqmotdoublet} in
${\cal L}^{(1)}= {\cal L}^{(1)}_{det}+ {\cal L}^{(1)}_{match}$,
the resulting Lagrangian is
\begin{eqnarray}
{\cal L}^{(1)}
&=&\frac{1}{m^2} \frac{1}{(4\pi)^2} \left(-\frac{\beta^3}{6}
(\varphi^\dagger\varphi)^3+ \frac{\beta^2}{12}
\partial_\mu (\varphi^\dagger\varphi)\partial^\mu (\varphi^\dagger\varphi)
+\frac{g'^2\beta}{12} (\varphi^\dagger\varphi) B_{\mu \nu} B^{\mu \nu}
\right.\label{afterem1}\\
&&\left.-\frac{g'^4}{60} \left( J^\mu_B J_{B\mu}+
i (\varphi^\dagger \lrover{D}_\mu \varphi) J^\mu_B
-\frac{1}{4}(\varphi^\dagger \lrover{D}^\mu \varphi)
(\varphi^\dagger \lrover{D}_\mu \varphi)\right)
\right)\label{afterem2}\\
&&-\frac{2}{3} \left(\delte+\frac{5}{3}\right)
\frac{g'^2}{m^2}(J^\mu_B+\frac{i}{2} (\varphi^\dagger \lrover{D}^\mu \varphi))
(\overline{\ell} F\gamma_\mu \ell)\label{afterem3}\\
&&
-\frac{2}{3}\left(\delte+\frac{4}{3}\right)
 \frac{g^2}{m^2}(\vec{J}^\mu_W+
 \frac{i}{2} (\varphi^\dagger \lrover{D}^\mu \vec{\tau}\varphi))
 (\overline{\ell} F\gamma_\mu \vec{\tau} \ell) \label{afterem4}\\
&&-\frac{g'}{3m^2} B^{\mu \nu}
(\overline{\ell} F Y_e \sigma_{\mu\nu} e\varphi+ \mrm{h.c.})
-\frac{g}{6m^2}  \vec{W}^{\mu \nu}
 (\overline{\ell} F Y_e\sigma_{\mu\nu} \vec{\tau} e\varphi+\mrm{h.c.})
 \hspace{2cm}\label{afterem5}\\
 && +\left(\frac{\beta}{m^2} (\varphi^\dagger \varphi)
(\overline{\ell} F Y_e e \varphi) + \mrm{h.c.} \right) \label{afterem6}\\
&&-\frac{(4\pi)^2}{m^2}  (\overline{\ell} F \gamma_\mu \ell)
(\overline{\ell} F \gamma^\mu \ell)\ .\label{afterem7}
\end{eqnarray}
The terms from \eq{totalefflagr} containing $\db^3$ and $\hat{F}$ have
been neglected as they are suppressed by two
or more leptonic Yukawa couplings.
 Again all couplings must be understood as bare effective Lagrangian
 couplings (barred couplings). We can write
 eqs.~(\ref{afterem1})--(\ref{afterem7}) in terms of
 $\overline{MS}$ renormalized couplings by just dropping the divergent
 contributions ($\delte \rightarrow 2\log(\mu/m)$).

It is clear that after using the equations of motion one can easily see the
plethora of
interesting processes that a single term in \eq{totalefflagr} can generate.

We will first study  how our effective Lagrangian can modify the
standard model mechanism for  SSB and what its  effects are on the
spectrum of physical gauge bosons.

The first term in (\ref{afterem1}) modifies the standard model
Higgs potential
and produces a shift in the vacuum expectation
value (VEV).  Since this is a one-loop effect one
should also use the full one-loop standard model effective potential to
analyse this shift. This gives  a global redefinition of the VEV and
has only consequences in the Higgs sector of the theory: the
ratio of the Higgs to the $W$ mass and the Higgs couplings are
changed, but there are no other effects. As we are not interested
in Higgs physics here we are not going to compute this shift.

The Higgs dependent terms of the effective Lagrangian can have
interesting effects when the Higgs develops a  VEV:
 \begin{equation}
 \label{vev}
\vev{\varphi} = \left(\begin{array}{c} 0 \\ v \end{array} \right)\ .
\end{equation}
 In what follows we will substitute the Higgs field by its VEV
and will neglect all the terms containing physical Higgses.

The second term in \eq{afterem1}, after SSB, produces a wave function
renormalization of the Higgs scalar. Again, since we are not interested in
Higgs interactions here we are going to neglect it.

The third term clearly gives a wave function renormalization of the
 $B_\mu$ gauge boson. Then,  one has to diagonalize simultaneously the
 kinetic term and the mass terms for the gauge bosons. After SSB
 we get the following kinetic term for the $B_\mu$ gauge boson
 \begin{equation}
 -\frac{1}{4} (1-\delta_m) B_{\mu\nu} B^{\mu\nu}\ ,
 \end{equation}
 where
  \begin{equation}
 \delta_m \equiv \frac{g'^2 \beta}{3(4\pi)^2} \frac{v^2}{m^2}\ .
  \end{equation}
 The $W$ gauge boson kinetic term and the mass terms remain unchanged.
 Then we can recover the canonical kinetic term by redefining the $B_\mu$
 field as follows
 \begin{equation}
  B_\mu \rightarrow (1-\delta_m)^{-1/2} B_\mu\ ;
  \end{equation}
  this will change the mass matrix of the gauge bosons. However as
  the $B_\mu$ field comes always with $g'$ we can recover the standard
  model form by redefining $g'$ as well
  \begin{equation}
  \label{hatg}
  \hat{g}' \equiv (1-\delta_m)^{-1/2} g'\ .
  \end{equation}

 It is not difficult to see that after SSB the last term in
 \eq{afterem2} gives
   \begin{equation}
  \label{ncoperators}
 \frac{1}{4} v^2 \frac{\hat{g}'^4}{60(4\pi)^2}\frac{v^2}{m^2}
(g W^{3\mu}-\hat{g}' B^\mu) (g W^3_\mu-\hat{g}' B_\mu) \ ,
  \end{equation}
 where we have already included the redefinition of $g'$ from \eq{hatg}.
  This term is an additional contribution to the neutral gauge bosons
  mass term. Adding it to the standard model contribution and
  diagonalizing the full gauge boson mass matrix, we find that the
  $W^3_\mu$ and $B_\mu$ gauge bosons can be expressed in
  terms of the physical photon $A_\mu$,and the $Z$-boson $Z_\mu$ as
  follows:
  \begin{eqnarray}
  \label{physbosons}
  W^3_\mu &=& \hat{s}_W A_\mu + \hat{c}_W Z_\mu  \\ \nonumber
  B_\mu &=& \hat{c}_W A_\mu - \hat{s}_W Z_\mu\ ,
  \end{eqnarray}
  where
  \begin{eqnarray}
  \hat{s}_W &=& \frac{\hat{g}'}{\sqrt{g^2+\hat{g}'^2}} \approx
  s_W \left(1+\frac{1}{2} c^2_W \delta_m\right) \\ \nonumber
  \hat{c}_W &=& \frac{g}{\sqrt{g^2+\hat{g}'^2}} \approx
  c_W \left(1-\frac{1}{2} s^2_W \delta_m\right)
  \end{eqnarray}
  and
  \begin{equation}
  s_W = \frac{g'}{\sqrt{g^2+g'^2}}
 \end{equation}
 is the tree-level weak mixing angle in the pure standard model.
 Similarly the relation between the electric charge and the gauge
 coupling is modified in comparison with the standard model one:
 \begin{equation}
 \label{echarge}
 e = \hat{g}' \hat{c}_W=g \hat{s}_W \approx
 g s_W \left(1-\frac{1}{2} c^2_W \delta_m\right)
  \ .
  \end{equation}
 With the same notation the physical masses are
 \begin{eqnarray}
 \label{physmasses}
 m^2_W &=& \frac{1}{2} g^2 v^2  \nonumber\\
 m^2_Z & = & \frac{m^2_W}{\hat{c}^2_W} (1+\delta_Z) \approx
 \frac{m^2_W}{c^2_W} (1+s^2_W \delta_m+\delta_Z)\ ,
 \end{eqnarray}
 where
 \begin{equation}
 \label{dz}
 \delta_Z = \frac{\hat{g}'^4}{60(4\pi)^2} \frac{v^2}{m^2}
 \end{equation}
 comes from \eq{ncoperators}.

  From these equations it is clear that only the $\delta_Z$ correction,
  which produces a relative shift form the standard relation between
  the masses of the $Z$ and the $W$ could in principle be observed;
  however, it is very small. All the other
  shifts related to $\delta_m$ can be absorbed in  the definition
  of the coupling $\hat{g}'$ or, equivalently,  in the weak mixing angle
  $\hat{s}_W$
  and then are not observable.

    Now we can go back to eqs.~(\ref{afterem1})--(\ref{afterem7})
  and write them in terms of the physical gauge bosons. The result
  we have found (neglecting all Higgs interactions) is:

  \begin{eqnarray}
{\cal L}^{(1)}
&=&-\frac{g^2}{2 m_W^2} \delta_Z
(c_W^2 J^\mu_A -J^\mu_Z)(c_W^2 J_{A\mu} -J_{Z\mu})
+ \frac{g}{\hat{c}_W} \delta_Z Z_\mu (c_W^2 J^\mu_A -J^\mu_Z)
 \label{current-current}\\
  &&+\frac{2}{3 } \frac{g}{m^2\hat{c}_W}
  \left(-(1-2\hat{s}_W^2) \left(\delte+\frac{4}{3}\right)+
\hat{s}_W^2\frac{1}{3}\right)
  \left(M_Z^2 Z^\mu+ \frac{g}{\hat{c}_W} J^\mu_Z\right)
   (\overline{\nu_L} F\gamma_\mu \nu_L)\hspace{1cm}\label{fczncoupling}\\
&&+\frac{2}{3} \frac{g}{m^2\hat{c}_W}
 \left(\left(\delte+\frac{4}{3}\right)+\hat{s}_W^2\frac{1}{3}\right)
\left(M_Z^2 Z^\mu+ \frac{g}{\hat{c}_W} J^\mu_Z\right)
 (\overline{e_L} F\gamma_\mu e_L)\label{fczecoupling}\\
&&
-\frac{2}{3}\left(\delte+\frac{4}{3}\right)
 \frac{g}{m^2} \left((\sqrt{2} M_W^2 W^+_\mu+J_\mu^\dagger)
  (\overline{\nu_L} F\gamma^\mu e_L)+\mrm{h.c.}\right)\label{fcwcoupling} \\
&&
-\frac{2}{9} \frac{e^2}{m^2} J_A^\mu (\overline{e} F\gamma_\mu e)\
-\frac{2}{3} \frac{e^2}{m^2} \left(\delte+\frac{5}{3}\right) J_A^\mu
(\overline{\nu_L} F\gamma_\mu \nu_L) \label{emfccoupling}\\
&& -\frac{1}{6} \frac{e}{m^2} A^{\mu\nu}
\left((\overline{e_L} F M_e \sigma_{\mu\nu} e_R) +
\mrm{h.c.}\right)\label{muegamma}\\
 &&+\frac{1}{6}\frac{g}{m^2\hat{c}_W}
 (1+\hat{s}^2_W) Z^{\mu\nu}
\left((\overline{e_L} F M_e \sigma_{\mu\nu} e_R)+\mrm{h.c.}\right)
\label{extraz}\\
&&-\frac{1}{3\sqrt{2}} \frac{g}{m^2}
\left(W^+_{\mu\nu}
(\overline{\nu_L} F M_e \sigma_{\mu\nu} e_R)+\mrm{h.c.}\right)
\label{extraw}\\
&&-\frac{(4\pi)^2}{m^2}  \left(
(\overline{e_L} F \gamma^\mu e_L) (\overline{e_L} F \gamma_\mu e_L)
+(\overline{\nu_L} F \gamma^\mu \nu_L)
(\overline{\nu_L} F \gamma_\mu \nu_L) \right.\nonumber
\\
&&
\hspace*{1.5cm}
+\left.
2(\overline{e_L} F \gamma^\mu e_L) (\overline{\nu_L} F \gamma_\mu \nu_L)
\right)
 \label{extrabox}
 \end{eqnarray}
 Here $M_e = Y_e v$ is the charged lepton mass matrix and
 $A^{\mu\nu} = \partial^\mu A^\nu-\partial^\nu A^\mu$ and
  $Z^{\mu\nu} = \partial^\mu Z^\nu-\partial^\nu Z^\mu$  are the
 field strengths of the photon and the $Z$ gauge boson, respectively.
 $\nu_L$ and $e_L$ are the components of the left-handed lepton
 doublet and $e_R$ are the right-handed parts of the charged
leptons.
 We have also defined the following neutral currents:
 \begin{eqnarray}
 J_A^\mu &=& \sum_f Q_f \overline{f} \gamma^\mu f\nonumber\\
 J_Z^\mu &=& \sum_f \overline{f} (v_f-a_f \gamma_5)\gamma^\mu f
 \end{eqnarray}
 where
 \begin{equation}
 v_f = \frac{1}{2}T_f^3-s_W^2 Q_f, \bla a_f = -\frac{1}{2} T_f^3
 \end{equation}
 and $J_\mu^\dagger$ is the usual charged current.
 Couplings with more than one gauge boson have not been included,
 and the term \rfn{afterem6} has been removed because after SSB
 it can be absorbed in a renormalization of the charged lepton
 mass matrix.

 As we have seen,  by using the equations of motion we have obtained a
 Lagrangian that displays  its physical content in a very transparent way.
 However,
 as stressed before, the use of the equations of motion is not necessary.
 We could have
 started with the unbroken Lagrangian, eqs.~\rfn{totalefflagr} and \rfn{lagr1},
 and then let the  Higgs field
 acquire a VEV without using the equations of motion at all.
 After diagonalization of the
 gauge boson mass matrices we would have obtained a quite different
 effective Lagrangian from that in
 eqs.~\rfn{current-current}--\rfn{extrabox}. However,  both Lagrangians
 give the same physics (at least at the level of precision  we are working
 here).  For example, the penguin operators in ${\cal L}^{(1)}$  (that is the
 first two operators in  the Lagrangian of \eq{totalefflagr}) when
 written in terms of the physical fields,
 give the following  interactions between $Z$-bosons and leptons:
\begin{eqnarray}
\label{zpenguin}
&&-\frac{g}{m^2\hat{c}_W}\frac{2}{3} \left(-(1-2 s_W^2)
\left(\delte+\frac{4}{3}\right)
+\frac{1}{3} s_W^2\right) \partial_\mu Z^{\mu\nu}
(\overline{\nu_L} F \gamma_\nu \nu_L)\nonumber\\
&&-\frac{g}{m^2\hat{c}_W}\frac{2}{3} \left(\left(\delte+\frac{4}{3}\right)
+\frac{1}{3} s_W^2\right) \partial_\mu Z^{\mu\nu}
(\overline{e_L} F \gamma_\nu e_L)\ .
\end{eqnarray}
 Clearly, from this interaction and for on-the-mass-shell
particles, we get the same amplitude for the $Z$ decay into a  pair
of different leptons as that obtained from
eqs.~(\ref{fczncoupling}) and (\ref{fczecoupling}).
 From \eq{zpenguin} we see that $Z$-exchange processes at low-energy,
 $q^2 \ll M_Z^2$,   are suppressed by $q^2/M_Z^2$. We get exactly the
 same answer from the compensation of the contributions from the
 two terms (the $Z^\mu$ and the $J_Z^\mu$ current terms) in \eq{fczncoupling} .

\mysection{Phenomenological consequences}
\label{pheno}

The terms in ${\cal L}^{(1)}$ give many interesting effects, apart from the
shifts in the gauge-boson masses, which have already been commented on.
For example, \eq{current-current} gives extra contributions to
flavour-conserving neutral-current processes.
Equation \rfn{emfccoupling} gives rise to four-fermion processes with
non-conservation of generational lepton numbers; for example they give
rise to
$\mu^- \rightarrow e^- e^- e^+$ and similar processes. To compute
these one has to take  into account also the contributions coming from
one-loop diagrams, with one insertion of the tree-level four-lepton
interaction.   The couplings in \eq{fczecoupling} allow the decay
of the $Z$ into different charged leptons. Equation (\ref{muegamma})
leads to $\mu \rightarrow e \gamma$ and similar processes. The amplitude
for this process
can easily be obtained from \eq{muegamma}.  Taking into account that,
without loss of
generality, $M_e$ can be taken diagonal with diagonal elements $m_a$ it is
given by
\begin{equation}
\label{ampmuegamma}
T(e_b \rightarrow e_a\, \gamma) = -i \frac{e}{3} F_{ab}
\bar{u}(p_a) \sigma_{\mu\nu} q^\nu (m_b R+m_a L) u(p_b) \epsilon^\mu(q)
\ ;
\end{equation}
$L$ and $R$ are, respectively,
the left-handed and right-handed chirality operators.
The amplitude \eq{ampmuegamma}
 is in complete agreement with the results obtained in
refs.~\cite{Pet82,LTV85,BS88}.

We are not going to study in detail the phenomenology of this model here.
An analysis of this  based on our effective
Lagrangian, will be presented elsewhere. However,
to give a flavour of the applicability of  our effective theory, and to
show how  one can work with it, we will comment on the calculation of
the decay $Z \rightarrow e^+_a\, e^-_b$ with both the full and the effective
theories.

In the effective theory we have two amplitudes, one due to the loop
diagram of fig.~4.c and the other coming from \eq{fczecoupling}.
As expected, even though
each of them contains a divergent contribution $\delte$,
 they give a finite answer when summed. The total amplitude is given
in appendix~\ref{fcdecay}, \eq{efffcamp}.

We also have performed a complete calculation of the amplitude
of this process in the full theory. After taking into  account the
diagrams with
self-energy insertions to the external fermion legs, the final answer is
finite.
The amplitude is given in \eq{fullamp} in terms of the functions
$f_1(w)$ and $f_2(w)$,  as explained in appendix~\ref{fcdecay}.
In the limit of $m \gg M_Z$  (that is,  $w \rightarrow 0$), the results
obtained in the full and in the effective theories are identical,
as they should be. In fact all matching conditions are designed with just
this in view.

{}From the full model amplitude of \eq{fullamp}, we obtain the following
branching ratio of the flavour-violating decay width ($a\not=b$)
relative to the standard model flavour-conserving one,
\begin{equation}
BR(Z\rightarrow \overline{e}_a e_b)=
\frac{\Gamma(Z\rightarrow \overline{e}_a e_b)}
{\Gamma(Z\rightarrow \overline{e}_a e_a)}= \left|F_{ab}\right| ^2
\frac{8\left| f_1(w)+\hat{s}^2_W f_2(w)\right|^2}{1+(1-4 \hat{s}_W^2)^2}\ .
\end{equation}
The branching ratio in the effective Lagrangian is given by the same
expression, but with $f_1(w)$ and $f_2(w)$ given in \eq{limfes}.
In   fig.~7,  we present this branching ratio in both
the full and the effective theories as a function of $m$.
Charged scalar Yukawa couplings, $ (f^\dagger f)_{ab}$, are taken to be
equal to 1.
{}From the figure it is clear that for masses of the scalar $h$
larger than the mass of the $Z$ boson the effective theory
gives a good  approximation.

\mysection{Discussion and conclusions}
\label{conclusion}

In this paper we have studied some questions related to the construction
and the use of effective Lagrangians,  by considering an
extension of the standard model that includes  a heavy  charged scalar
 coupled to the leptonic doublet.

 Starting from the full renormalizable model, we have built a low-energy
effective field theory by integrating out the heavy scalar. This was done
at the one-loop
 level and keeping only
 operators of dimension six or less.

 Functional methods were used to obtain, in a compact and gauge-invariant
 form, all operators generated by tree-level and one-loop diagrams
 containing only heavy scalars in the internal lines. The result
 includes only diagrams which are one-particle-irreducible with respect
 to the light particles.
 In appendix~\ref{det} we give
 a detailed explanation of this calculation.
 This is not the complete answer for the
 effective Lagrangian, since there are many one-loop diagrams
 in the full theory,
 with heavy and light particles in the loop that are not completely
 taken into account by one-loop diagrams involving tree-level
 effective couplings. To obtain these additional contributions we
 have  calculated several Green functions in both the
 full and the effective theories and required that the results of
 both calculations match for  small momenta compared with the
 heavy scalar mass. The result can be expressed as a linear combination
 of gauge-invariant operators, which must be included in the effective
 Lagrangian.

 Adding both contributions, from the operators
 corresponding to the diagrams with only
 heavy lines and from those with heavy and light internal particles,
  we  have obtained the complete bare one-loop effective Lagrangian,
 including operators up to dimension six.
 It contains several operators with infinite
 coefficients (dimensional continuation was employed to regularize
  all divergent integrals, then UV divergences appear as poles in $1/(D-4)$).
  By using
 an $\overline{MS}$ scheme to renormalize both the full and the
 effective theories, we obtained the matching conditions for the
 running couplings, which express the renormalized couplings
 of the effective theory in terms of the renormalized couplings of
 the full model. To avoid large logarithms, these matching conditions should
be evaluated at a scale around the heavy scalar mass and then the
renormalization group used to bring the couplings to any other lower scale.

 We used this simple example to discuss the behaviour of the couplings
of a generic effective Lagrangian. In general all the operators in the
effective Lagrangian with the same quantum numbers mix under the
renormalization group. The effective couplings obey a standard
renormalization group first-order differential equation controlled
by the anomalous-dimension matrix. It is clear that the only information
that can be extracted from the effective Lagrangian, without knowing the
full theory, is this
anomalous-dimension matrix. However, to obtain the couplings it is
necessary to solve the renormalization group equations, and this requires
the knowledge of $n$ boundary conditions
(if $n$-operators are involved in the mixing).
One-loop calculations with the effective Lagrangian  can
only be used to relate the couplings of operators at different scales, but give
 no
information at all on the actual value of those couplings. Only experiment,
or a more complete theory, can give new information on the  values
of the effective Lagrangian couplings. This trivial observation is of
importance when using the effective Lagrangian approach to classify
the sort of  new physics one could find in future experiments.

After renormalization we obtain an effective Lagrangian that can
be split in three pieces.
The first one,  ${\cal L}_{SM}$, is just the standard model
Lagrangian
(expressed in terms of effective couplings); the second one is the
four-lepton interaction obtained from the full theory at tree level,
 ${\cal L}^{(0)}$; the third piece, ${\cal L}^{(1)}$ contains all
dimension-six operators generated at the one-loop level in the full model.

This effective Lagrangian
was rewritten by using
the standard model classical equations of motion
 in order to display  its physical content more transparently.
Some caution is needed when using the equations of motion,
especially for operators that are going to be inserted in
loop diagrams.
In our
case, however,  the equations of motion are used only in
operators that are generated at one loop in the full theory and that  are
supposed to be used only at tree level in the effective Lagrangian. Then,
the use of equations of motion is completely legitimate.
 We used them for the unbroken
theory, but it is worth while to stress that the equations of
motion might also be used  after SSB (or might not be used at all) and
this would not change any physical result.

 We also discussed the consequences of SSB on our effective Lagrangian by
 substituting the VEV of the doublet and neglecting all Higgs interactions.
 Apart from a negligible  modification of the relation between the
masses of the vector bosons, the most interesting consequence
of the model is due to the different operators contributing to processes
with violation of the generational lepton numbers.
We made a short review of some processes
generated by our one-loop effective Lagrangian, $e_a^- \rightarrow e_b^-
\gamma$, $e_a^- \rightarrow e_b^- e_c^+ e_c^+$,  etc (where
$a,b,c$ are different flavour indices). A more
detailed phenomenological analysis will be presented elsewhere.

Finally,
to see how one can use our effective Lagrangian, we have calculated
the decay width of the $Z$ gauge boson to a pair of different leptons.
It contains contributions from one-loop diagrams with one insertion
of the tree-level four-fermion operator and direct contributions from
operators generated at the one-loop level. We have done also the
calculation in the full theory and compared the results. For  a  mass
of the charged scalar $m$ larger than the $Z$ mass, the effective theory
calculation gives a good approximation.

\section*{Acknowledgements}

We thank J. Bernab\'eu, A. Casas, S. Fanchiotti, S. Peris and A. Pich
for helpful discussions
on the subject of this paper and for a critical reading
of the manuscript.
This work was  supported in part by
CICYT, Spain, under grant AEN93-0234.
\vfill\eject
\appendix

\begin{center}{\Large\bf APPENDICES}\end{center}

\mysection{The determinant of the fluctuation operator}
\label{det}
In this appendix we calculate the effective Lagrangian
contribution coming from the determinant of a  fluctuation operator
of the form $O = (-D^2-m^2-U(x))$. It is given by
\begin{equation}
\label{ap1}
i \int d^4 x {\cal L}^{(1)}(x)=
\log{\det{(O)}^{-1}}=
-\Tr{\log(O)}=
-\Tr{\log (-D^2-m^2-U(x))} \ ,
\end{equation}
where $D_\mu$ is the covariant derivative for a generic gauge group,
for example an $SU(N)$, and $U(x)$ is a general matrix valued function
of $x$.

Powerful methods to calculate traces of these kinds of operators
have been developed
in the last years \cite{Che88,AF85,Cha85,Zuk85,Zuk86,Cha86,Gai86,Bal89}.
We will employ some techniques developed in all of those papers, but we will
follow more closely those in  ref.~\cite{Bal89}.

We  will use the following notation
\begin{equation}
D_\mu = \partial_\mu + A_\mu(x) , \bla  A_\mu(x) = -i g T^a A^a_\mu(x) \ ,
\end{equation}
 where $T^a$ are the generators of the gauge group. They satisfy the
normalization condition $\tr{T^a T^b} = \frac{1}{2} \delta^{ab}$. Then,
the covariant derivative acts on $U(x)$ as follows:
\begin{equation}
D_\mu U(x) = \partial_\mu U(x)+ [A_\mu(x), U(x)] \ ,
\end{equation}
The field strength tensor is defined as
\begin{equation}
F_{\mu\nu} = [ D_\mu, D_\nu] \ .
\end{equation}
We will use the metric $(+,-,-,-)$  and the following
conventions for the momentum operator and plane wave states:
\begin{equation}
\hat{p}_\mu = i\partial_\mu,\bla   \langle x | p \rangle = e^{-ipx}\ .
\end{equation}
 The normalization of the states is  (in $D$ dimensions)
\begin{equation}
\int_x \ket{x}\bra{x} = 1,\bla  \int_p \ket{p}\bra{p} =1
\end{equation}
 and $\int_x = \int d^D x$ and $\int_p = \int d^D p/(2\pi)^4$.

Let  $O$ be an operator; then we understand the trace of that operator
to be
\begin{equation}
\Tr{O} = \int_x \tr{\bra{x} O \ket{x}} = \int_p \tr{\bra{p} O \ket{p}}\ ,
\end{equation}
 where ``Tr'' means trace over all degrees of freedom, space-time and
internal, while ``tr" is the trace  over only the internal degrees of freedom.
Then
\begin{equation}
\label{trO}
\Tr{O} = \int_x \int_p \tr{e^{ipx} \vec{O}_x e^{-ipx}}\ .
\end{equation}
 Here $O_x$ is the operator $O$ in the representation of positions
$\bra{x} O \ket{\phi} = \vec{O}_x \langle x | \phi\rangle = \vec{O}_x \phi(x)$.
In our case we have
\begin{equation}
O= \log(\Pi^2-m^2-U),\bla \mrm{where}\bla
\Pi_\mu = i D_\mu = \hat{p}_\mu + i A_\mu\ .
\end{equation}
 By using \eq{trO} and the operator identity
 $e^{ipx} f(\Pi) e^{-ipx} = f(\Pi+p)$, we find
\begin{equation}
i \Tr{\log(\Pi^2-m^2-U)} =
i\int_x\int_p \tr{\log(p^2-m^2+2p\Pi+\Pi^2-U) } \mathbold{1}\ ,
\end{equation}
 where the factor $\mathbold{1}$ at the end indicates that the operators
 act on the identity.
Now we can expand the logarithm in the following form
\begin{equation}
i\int_x \int_p \tr{\log(p^2-m^2)-\sum_{n=1}^\infty \frac{(-1)^n}{n}
\frac{(2p\Pi+\Pi^2-U)^n}{(p^2-m^2)^n} } \mathbold{1} \ .
\label{effprel}
\end{equation}
 The first term is the usual Coleman-Weinberg term. It is a constant term,
which only contributes to the energy density, and we will drop it.
Comparing \eq{effprel} with the expression for the effective Lagrangian,
\eq{ap1},
 we find
\begin{equation}
\label{lexpansion1}
{\cal L}^{(1)} = \sum_{n=1}^\infty \frac{(-1)^{n+1}}{n}
\tr{i\int_p \frac{(2p\Pi+\Pi^2-U)^n}{(p^2-m^2)^n} } \mathbold{1}\ .
\end{equation}
 It is clear that \eq{lexpansion1} gives an expansion in powers of $1/m^2$
\begin{equation}
\label{lexpansion2}
{\cal L}^{(1)} = \sum_{n=1}^\infty \frac{c_n}{m^{2n-4}} O_n\ ,
\end{equation}
 where $O_n$ are traces of gauge-invariant operators of dimension $2n$ built
from $A_\mu$ and $U$ (and their covariant derivatives). In general
\begin{equation}
\label{lop1}
O_n = \sum_i \gamma_i^{(n)} \tilde{O}^{(i)}_n \ .
\end{equation}
Here $\tilde{O}^{(i)}_n$ is a linearly independent set of traces of operators
of
dimension $2n$ (operators that can be related using partial integration
of $D_\mu$, cyclic permutations inside the trace, or Bianchi
identities are not considered to be linearly independent).
We use the following basis for these traces of gauge-invariant operators:
\begin{eqnarray}
\tilde{O}_1 &= & (\tr{U}) \nonumber \\
\tilde{O}_2 &= & (\tr{U^2},\ \tr{F_{\mu\nu} F^{\mu\nu} }) \nonumber \\
\tilde{O}_3 &= & (\tr{U^3},\ \tr{(D_\mu U)^2},\
\tr{F_{\mu\nu} U F^{\mu\nu} },\ \tr{D_\mu F^{\mu\nu} D^\sigma
F_{\sigma\nu} },\
\tr{F_{\mu\nu} F^{\nu\sigma} F_\sigma^{\ \mu} })\ .\nonumber
\end{eqnarray}

The normalization of the $O_n$ is such that
 \begin{equation}
\label{coeff1}
c_n = \left(\frac{m^2}{4\pi\mu^2}\right)^{D/2-2} \frac{1}{(4\pi)^2}
\Gamma(n-D/2)\ .
\end{equation}
 All the integrals will be  dimensionally regularized
and $\mu$ is the dimensional-regularization scale.
As we will see, with this normalization the coefficients $\gamma_i^{(n)}$
will be just numbers and independent from $D$. Then, our task is to
evaluate the coefficients $\gamma_i^{(n)}$.
They could be evaluated directly by expanding \eq{lexpansion1}, performing
the momentum integrals and using integration by parts to rewrite it in a
canonical form. However, it is better to use another method
\cite{Bal89,Hoo73}. The trick is as follows: as we know that
 eqs.~\rfn{lexpansion1},
 \rfn{lexpansion2} and \rfn{lop1} are valid for any
 $A_\mu$ and any $U$, with the $\gamma_i^{(n)}$ independent from $A_\mu$
 and $U$, we can compute $\gamma_i^{(n)}$ using a particular configuration
 of $A_\mu$ and $U$.  Then, the resulting $\gamma_i^{(n)}$ will be valid for
 any $A_\mu$ and $U$. As we want to compute \rfn{lexpansion1} and we have
 $2p\Pi+\Pi^2-U= 2ip A+2p\hat{p}+\hat{p}^2+i(\hat{p} A+A\hat{p})-
 A^2-U$, it is obvious that if we choose $\partial_\mu A_\nu =0$ and
 $U=-A^2$ we will have $(2p\Pi+\Pi^2-U)^n \mathbold{1} = (2ip A)^n$. Thus,
 to calculate \eq{lexpansion1} we will choose the following (constant)
 configuration for the field $A_\mu$:
 \begin{equation}
 N_\mu \equiv A_\mu \bla\mrm{such as}\bla \partial_\nu A_\mu = 0
\bla \mrm{and}\bla U=-A^2 \ .
\end{equation}
To avoid confusion we will denote the field $A_\mu$ as $N_\mu$
in this special configuration. In this  configuration we obviously have
\begin{equation}
\label{specialconf}
U=-N^2,\bla F_{\mu\nu} = [ N_\mu, N_\nu],\bla D_\mu G = [N_\mu,G]\ ,
\end{equation}
where $G$ is any matrix valued function of $A_\mu$ and $U$ in the special
configuration (hence, constant with respect to space-time variables). The
important point is that  \eq{specialconf} can be
inverted, thus allowing us to pass from the special configuration to the
general configuration.

Since in the special configuration \rfn{specialconf} we have
\[
(2p\Pi+\Pi^2-U)^n\ \mathbold{1} = (2ip N)^n
\]
\eq{lexpansion1} reads
\begin{eqnarray}
\label{lagconspe1}
{\cal L}^{(1)} &=& \sum_{n=1}^\infty \frac{(-1)^{n+1}}{n} \tr{i\int_p
\frac{(2ipN)^n}{(p^2-m^2)^n}} \\ \nonumber
& =& \sum_{n=1}^\infty \frac{(-1)^{n+1}4^n}{2n} i\int_p
\frac{p_{\mu_1}p_{\mu_2}\cdots p_{\mu_{2n}}}{(p^2-m^2)^{2n}}
\tr{N^{\mu_1}N^{\mu_2}\cdots N^{\mu_{2n}}}\ ,
\end{eqnarray}
where we have taken into account that integrals with an odd number of
$p$'s cancel under symmetric integration and we have redefined
$n\rightarrow 2n$. Now we can use that
\begin{equation}
\label{integsym}
i\int_p
\frac{p^{\mu_1}p^{\mu_2}\cdots p^{\mu_{2n}}}{(p^2-m^2)^{2n}}
= (-1)^{n+1} \left(\frac{m^2}{4\pi\mu^2}\right)^{D/2-2}
\frac{m^4}{(4\pi)^2} \frac{\Gamma(n-D/2)}{m^{2n} 2^n \Gamma(2n)}
S_n^{\mu_1  \mu_2 \cdots \mu_{2n}}\ .
\end{equation}
Here $S_n^{\mu_1  \mu_2 \cdots \mu_{2n}}$ is the completely symmetric
tensor with $2n$ indices built only with the metric tensor $g_{\mu\nu}$.
For instance $S_1^{\mu_1\mu_2} = g^{\mu_1\mu_2}$ and
$S_2^{\mu_1\mu_2 \mu_3 \mu_4}= g^{\mu_1\mu_2} g^{\mu_3\mu_4}+
g^{\mu_1\mu_3}g^{\mu_2\mu_4}+ g^{\mu_1\mu_4} g^{\mu_2\mu_3}$. In
general it contains $(2n-1)!! = (2n-1)(2n-3)\cdots 1$ terms. Inserting
\eq{integsym} in \eq{lagconspe1}, we finally obtain
\begin{equation}
\label{lagconspe2}
{\cal L}^{(1)} =
\sum_{n=1}^\infty
\left(\frac{m^2}{4\pi\mu^2}\right)^{D/2-2}
\frac{m^4}{(4\pi)^2} \frac{\Gamma(n-D/2)}{m^{2n}}\frac{2^n}{(2n)!}
S_n^{\mu_1  \mu_2 \cdots \mu_{2n}}
\tr{N^{\mu_1}N^{\mu_2}\cdots N^{\mu_{2n}}}\ .
\end{equation}
Comparing this result with \eq{lexpansion2} and \eq{coeff1}
we find that in the special
configuration (\ref{specialconf}) we have
\begin{equation}
O_n = \frac{2^n}{(2n)!} \tr{S_n(N)}\ ,
\end{equation}
where $S_n(N) \equiv S_n^{\mu_1  \mu_2 \cdots \mu_{2n}}
N_{\mu_1}N_{\mu_2}\cdots N_{\mu_{2n}}$ represents the sum of all
possible
products of $2n$ $N_\mu$'s contracted pairwise. For instance, $S_1(N) =
N^2$ and $S_2(N) = (N^2)^2 +N_\mu N_\nu N^\mu N^\nu+ N_\mu N^2 N^\mu$.
Traces of different operators can be related by using the cyclic property of
the trace. We find for example
$\tr{S_2(N)}= 2\tr{(N^2)^2} +\tr{N_\mu N_\nu N^\mu N^\nu}$.
 Then, in the special configuration we have
\begin{equation}
O_n = \sum_i \delta_i^{(n)} \hat{O}^{(i)}_n\ ,
\end{equation}
where $\hat{O}^{(i)}_n$ are traces of strings of $N_\mu$'s contracted
pairwise and the coefficients $\delta_i^{(n)}$ include the normalization
factor $2^n/(2n)!$.

If we define the set of linearly independent  traces built from $N_\mu$
as follows:
\begin{eqnarray}
\hat{O}_1 &= & (\tr{N^2}) \\ \nonumber
\hat{O}_2 &= & (\tr{(N_\mu N_\nu)^2},\tr{(N^2)^2}) \\ \nonumber
\hat{O}_3 &= & (\tr{(N_\mu N_\nu N_\sigma)^2},\tr{(N^2)^3},
\tr{(N^2 N_\mu)^2}, \tr{(N_\mu N_\nu N^\mu)^2},\tr{N^2(N_\mu N_\nu)^2})\ ,
\nonumber
\end{eqnarray}
we find
\begin{eqnarray}
O_1 & = & \hat{O}_1 \nonumber\\
O_2 & = & \frac{1}{6} \hat{O}_2^{(1)}+\frac{1}{3}\hat{O}_2^{(2)}
\nonumber\\
O_3 & = & \frac{1}{90} \hat{O}_3^{(1)}+\frac{1}{45} \hat{O}_3^{(2)}+
\frac{1}{30} \hat{O}_3^{(3)}+\frac{1}{30} \hat{O}_3^{(4)}+\frac{1}{15}
\hat{O}_3^{(5)}\ ,\nonumber
\end{eqnarray}
but we know from \eq{lop1} that in the general case $O_n$ can be expressed
as a linear combination of traces of gauge-invariant operators
$\tilde{O}^{(i)}_n$.
Then, in the special configuration we can  write $O_n$ in terms of the
two sets of operators
\begin{equation}
O_n = \sum_i \delta_i^{(n)} \hat{O}^{(i)}_n =
\sum_i \gamma_i^{(n)} \tilde{O}^{(i)}_n\ ,
\end{equation}
but in the special configuration we can relate the operators
$\tilde{O}^{(i)}_n$ with the operators $\hat{O}^{(i)}_n$ by using
\eq{specialconf}. In general we will have
\begin{equation}
\tilde{O}_n^{(i)} = \sum_j (P_n)_{ij} \hat{O}^{(j)}_n\bla \mrm{and}\bla
\delta_i^{(n)}  = \sum_j \gamma_j^{(n)} (P_n)_{ji}\ .
\end{equation}
The important point of these equations is that the above linear
transformation $P_n$ can can be inverted. This
will allow us to compute the $\gamma$'s in terms of the already known
$\delta$'s. In matrix form we obtain
\begin{equation}
\gamma^{(n)} = (P^T_n)^{-1} \delta^{(n)}\ ,
\end{equation}
where $P_n^T$ is the transposed matrix of $P_n$. We skip the details of
a rather tedious calculation and quote only the final results valid for an
arbitrary configuration
\begin{eqnarray}
\label{resdet}
O_1 &= & -\tr{U} \nonumber \\
O_2 &= & \frac{1}{2} \tr{U^2}+\frac{1}{12} \tr{F_{\mu\nu} F^{\mu\nu} } \\
O_3 &= & -\frac{1}{6} \tr{U^3}+\frac{1}{12}\tr{(D_\mu U)^2}
-\frac{1}{12}\tr{F_{\mu\nu} U F^{\mu\nu} } \nonumber \\
& &+\frac{1}{60}\tr{D_\mu F^{\mu\nu} D^\sigma F_{\sigma\nu} }
-\frac{1}{90}\tr{F_{\mu\nu} F^{\nu\sigma} F_\sigma^{\ \mu} }\ .\nonumber
\end{eqnarray}
These results agree with the results obtained in \cite{Bal89}
 by using proper time methods (after passing
to Minkowski space).

The effective Lagrangian obtained from the determinant of the
fluctuation operator is given by \eq{lexpansion2} with the $O_n$ given
above and
with the $c_n$ given in \eq{coeff1}. For $D=4-2\epsilon$ and in the limit
$\epsilon \rightarrow 0$ we obtain
\begin{eqnarray}
c_1 &=& -(\delte+1) \frac{1}{(4\pi)^2} \nonumber \\
c_2 &=& \delte \frac{1}{(4\pi)^2} \nonumber \\
c_3 &=& (n-3)! \frac{1}{(4\pi)^2}\bla n\geq 3\ ,
\end{eqnarray}
where
\begin{equation}
\label{deltae}
\delte = \frac{1}{\epsilon} -\gamma+\log(4\pi) +2\log(\mu/m)
\equiv \frac{1}{\hat{\epsilon}} +2\log(\mu/m)
\end{equation}
contains the divergent part when $D\rightarrow 4$ and only appears in the
first two coefficients.

\mysection{Calculation of processes with heavy-light particles in the loops}
\label{heavy-light}

Here we  collect the relevant amplitudes needed to obtain the effective
operators generated by one-loop diagrams in the full theory
with heavy-light particles in
the loops. We have always used dimensional regularization with
anticommuting $\gamma_5$.

\subsection{Lepton doublet self-energies}

{}From the self-energy diagrams in the full theory (fig.~3) we find
\begin{equation}
\label{selfe}
T_{(3)} =  \frac{(f^\dagger f)_{ab}}{(4\pi)^2}
\left(2\left(\delte + \frac{1}{2}\right)
+ \frac{2}{3} \frac{p^2}{m^2} \right) \bar{u}(p) \sla{p} L u(p) \ ,
\end{equation}
where $L=\frac{1}{2} (1-\gamma_5)$ is the left-handed chirality
projector.  In the effective theory, the doublet self-energy diagram with
the four-fermion coupling (fig.~3.b) is zero in dimensional regularization
because it is a massless tadpole-like diagram.

\subsection{Penguins}

The diagrams of figs.~4.a and 4.b give the following amplitudes for
the coupling of the $B_\mu$ to the lepton doublet
\begin{eqnarray}
T^B_{4a} &=&  F_{ab}\, \frac{g'}{m^2}\, \bar{u}(p_2)\left\{
m^2\left(\delte+\frac{1}{2}\right) \gamma_\mu
-\frac{2}{3} \left(\delte+\frac{4}{3}\right) (q^2
 \gamma_\mu-\sla{q}q_\mu)\right.\nonumber\\
&&+ \frac{1}{6} \left( (\sla{p}_1+\sla{p}_2)(p_1+p_2)_\mu
+(p_1^2+p_2^2)\gamma_\mu \right)\nonumber \\
&&\left. +\frac{i}{2} (\sla{p}_2 \sigma_{\mu\nu} q^\nu+
 \sigma_{\mu\nu} q^\nu \sla{p}_1) \right\}u(p_1) \epsilon^\mu(q) \nonumber\\
T^B_{4b} & = &  F_{ab}\,  \frac{g'}{m^2}\, \bar{u}(p_2)\left\{
 -m^2 2 \left(\delte+\frac{1}{2}\right) \gamma_\mu
-\frac{2}{9}  (q^2 \gamma_\mu-\sla{q}q_\mu)\right.\nonumber\\
&&\left.- \frac{1}{3} \left( (\sla{p}_1+\sla{p}_2)(p_1+p_2)_\mu
+(p_1^2+p_2^2)\gamma_\mu \right)\right\} u(p_1) \epsilon^\mu(q)\ .
\nonumber
\end{eqnarray}
Here $q=p_2-p_1$, and
 we have kept all external momenta off-the-mass-shell.  In the diagram~4.a
  we split the charged scalar propagator as described in
section~\ref{matching}. The result given here corresponds
to the difference between the full theory diagram (fig.~4.a) and the
vertex obtained in the effective theory with the four-fermion interaction
(fig.~4.c).

The full amplitude for the $B_\mu$ vertex is
\begin{eqnarray}
T^B &=&  F_{ab}\, \frac{g'}{m^2}\, \bar{u}(p_2)\left\{
-m^2 \left(\delte+\frac{1}{2}\right) \gamma_\mu
-\frac{2}{3} \left(\delte+\frac{5}{3}\right) (q^2
\gamma_\mu-\sla{q}q_\mu)\right. \nonumber \\
&&- \frac{1}{6} ( (\sla{p}_1+\sla{p}_2)(p_1+p_2)_\mu
+(p_1^2+p_2^2)\gamma_\mu )\nonumber\\
&&\left. +\frac{i}{2} (\sla{p}_2 \sigma_{\mu\nu} q^\nu+
 \sigma_{\mu\nu} q^\nu \sla{p}_1) \right\} u(p_1) \epsilon^\mu(q)
 \  .\label{ampB}
 \end{eqnarray}

For the vertex of the $\vec{W}_\mu$ only the diagram~4.a exists since
it does not couple to the $SU(2)$ singlet scalar. The result is
\begin{eqnarray}
T^W &=&  F_{ab}\, \frac{g}{m^2}\, \bar{u}(p_2) \left\{
 m^2 \left(\delte+\frac{1}{2}\right) \gamma_\mu
-\frac{2}{3} \left(\delte+\frac{4}{3}\right) (q^2
\gamma_\mu-\sla{q}q_\mu)\right. \nonumber\\
&&+ \frac{1}{6} \left( (\sla{p}_1+\sla{p}_2)(p_1+p_2)_\mu
+(p_1^2+p_2^2)\gamma_\mu \right)\nonumber\\
&& \left.+\frac{i}{2} (\sla{p}_2 \sigma_{\mu\nu} q^\nu+
 \sigma_{\mu\nu} q^\nu \sla{p}_1) \right\} \tau_i\, u(p_1) \epsilon_i^\mu(q)
 \ .\label{ampW}
\end{eqnarray}

{}From these amplitudes together with the self-energy amplitudes, one
can easily reconstruct the gauge-invariant operators that generate
them as done in section \ref{matching}.

\subsection{Seagull diagrams}

{}From the diagram of fig.~5.b we obtain the amplitude
\begin{equation}
\label{amp5a}
 \frac{\beta}{m^2} F_{ab} \bar{u}(p_2)(\sla{p}_1+\sla{p}_2) u(p_1)\ ,
\end{equation}
while for the diagram of fig.~5.a we get
\begin{equation}
\label{amp5b}
  \frac{1}{m^2}\hat{F}_{ab} \bar{u}(p_2)\left(2 (\delte+1)
(\sla{q}_1+\sla{q}_2)
- (\sla{p}_1+\sla{p}_2) \right) u(p_1)\ .
\end{equation}
Here $q_1$ and $q_2$ are the momenta of an incoming $\widetilde{\varphi}$
and an outgoing $\widetilde{\varphi}$ respectively,  and
$\hat{F}$ is defined in \eq{defhatF}.
As always,  we have used the splitting of the charged scalar
propagator $1/(k^2-m^2)= -1/m^2 + k^2/(m^2(k^2-m^2))$ and have
only computed the second term. The first term gives rise to the
effective Lagrangian contribution given by the diagram of fig.~5.c.

{}From \eq{amp5b} we obtain the following effective operators
\begin{equation}
(\delte+1) 2 \frac{i}{m^2}
(\overline{\ell} \gamma^\mu (D_\mu \widetilde{\varphi})
\hat{F} \widetilde{\varphi}^\dagger \ell)
-\frac{i}{m^2} (\overline{\ell} \widetilde{\varphi}\hat{F}
 \widetilde{\varphi}^\dagger \db \ell) +\mrm{h.c.}
\ ,
\end{equation}
which can be written in the form of \eq{op5b} after using the following
$SU(2)$ Fierz transformations:
\begin{eqnarray}
((D_\mu\widetilde{\varphi}) \widetilde{\varphi}^\dagger) &=&
\frac{1}{2} ((D_\mu\varphi)^\dagger \varphi) \mathbold{1}+
\frac{1}{2} ((D_\mu\varphi)^\dagger \vec{\tau} \varphi) \vec{\tau}\nonumber\\
(\widetilde{\varphi} \widetilde{\varphi}^\dagger) &=&
\frac{1}{2} (\varphi^\dagger \varphi) \mathbold{1}+
\frac{1}{2} (\varphi^\dagger \vec{\tau} \varphi) \vec{\tau}\nonumber\ .
\end{eqnarray}
 Here $\mathbold{1}$ is the $2\times 2$ identity matrix.

\subsection{Four-fermion interactions}

The amplitude corresponding to the box with the two charged scalars in
the loop (fig.~6.c and crossed) has the following form
\begin{equation}
\label{ampbox}
-\frac{(4\pi)^2}{m^2} F_{ab} F_{cd} (\bar{u}(p_1) \gamma_\mu u(p_2))
(\bar{u}(p_3) \gamma_\mu u(p_4) )\ .
\end{equation}

\mysection{Lepton-flavour-changing $Z$ decay}
\label{fcdecay}

In this appendix we give some results of the calculation of the decay width
for $Z\rightarrow \bar{e}_a e_b$ done with both the full and the effective
Lagrangians.

We show in fig.~4.c the diagram responsible for this decay in
the effective theory (the wavy line is, in this case, a $Z$ gauge boson,
the external leptons are charged leptons or neutrinos,  and the  lines
in the loop represent neutrinos or charged leptons, respectively).
The corresponding
amplitudes are (for massless external leptons and all particles
on-the-mass-shell):
\begin{equation}
\label{amp4c}
T(Z\rightarrow \overline{e}_a e_b)_{(4.c)} =
  \frac{g}{\hat{c}_W} F_{ab} \frac{2}{3} \frac{M_Z^2}{m^2}
\left(\log(M_Z^2/m^2)-i\pi -\delte-\frac{5}{3}\right)
\overline{u}(p_2) \sla{\epsilon}(q) L v(p_1)
\ .
\end{equation}
To this amplitude we should add the contribution coming from the operators
generated through matching, \eq{fczecoupling}. The total result is
\begin{equation}
T(Z\rightarrow \overline{e}_a e_b)_{eff} =
 \frac{g}{\hat{c}_W} F_{ab} \frac{2}{3} \frac{M_Z^2}{m^2}
\left(\log(M_Z^2/m^2)-i\pi-\frac{1}{3} +\hat{s}_W^2\frac{1}{3}\right)
 \overline{u}(p_2) \sla{\epsilon}(q) L v(p_1) \  . \label{efffcamp}
\end{equation}
For the $Z$ decay to neutrinos,  a similar calculation gives  the following
result:
\begin{eqnarray}
&&T(Z\rightarrow \overline{\nu}_a \nu_b)_{eff} = \nonumber\\
&&\frac{g}{\hat{c}_W} F_{ab} \frac{2}{3} \frac{M_Z^2}{m^2}
\left(-(1-2 \hat{s}_W^2) (\log(M_Z^2/m^2)-i\pi-\frac{1}{3})
+\hat{s}_W^2\frac{1}{3}\right)
 \overline{u}(p_2) \sla{\epsilon}(q) L v(p_1) \ . \hspace{1cm}
\end{eqnarray}
In both cases the divergent term coming from the loop integration
has been cancelled by the one-loop effective operator contribution.

Previous results  should be compared with the full Lagrangian amplitudes,
\begin{eqnarray}
\label{fullamp}
T(Z\rightarrow \overline{e}_a e_b)_{full} & =&
 \frac{g}{c_W}  F_{ab}
\left( -f_1(w)- s^2_W f_2(w)\right)
\bar{u}(p_2) \sla{\epsilon}(q) L v(p_1)  \\
T(Z\rightarrow \overline{\nu}_a \nu_b)_{full} &=&
 \frac{g}{c_W}  F_{ab}
\left( (1-2 s^2_W) f_1(w)-s^2_W f_2(w)\right)
\bar{u}(p_2) \sla{\epsilon}(q) L v(p_1) \ ,\hspace{1cm}
\end{eqnarray}
where,  using the variable
\begin{equation}
w= -\frac{M_Z^2}{m^2} - i\eta\bla (\eta \rightarrow 0^+)\ ,
\end{equation}
we have
\begin{equation}
f_1(w) = \frac{1}{2} + \frac{2}{w} - \frac{2+w}{w} \log (w)
 - \frac{2}{w^2} \log (1 - w)\log (w) - \frac{2}{w^2} \mrm{Li}_2(w)
\end{equation}
and
\begin{eqnarray}
f_2(w) &=& -5 - {4\over w} + \frac{8}{w^2}{\left( \mrm{Li}_2 \left({2\over
           {1 - {\sqrt{1 + {4\over w}}}}}\right) +
        \mrm{Li}_2 \left({2\over {1 + {\sqrt{1 + {4\over w}}}}}\right) \right)
}
\nonumber \\
&&+2\left(1+\frac{2}{w}\right){{\sqrt{1 + {4\over w}}}\,
      \log \left({{1 + {\sqrt{1 + {4\over w}}}}\over
         {-1 + {\sqrt{1 + {4\over w}}}}}\right)}
\ .
\end{eqnarray}
Here $Li_2(w)$ is the dilogarithmic function.
The functions $f_1(w)$ and $f_2(w)$  have the following asymptotic
values for $w \rightarrow 0$:
\begin{equation}
\label{limfes}
f_1(w)  \rightarrow  \frac{2}{3} w \left(\log(w)-\frac{1}{3}\right), \bla
f_2(w)  \rightarrow  \frac{2}{9} w\  .
\end{equation}
 These results are in complete agreement with the effective Lagrangian
calculation. In fact the logarithmic contribution coming from
$f_1(w)$ can easily be obtained from the calculation of diagram~4.c.
However,
the non-logarithmic part of that amplitude is arbitrary, in fact
divergent. The only way to fix it  is by matching the full theory.

\vfil\eject

\vfil\eject

\section*{Figure captions}

{\bf Figure 1:} Diagram a) is the tree-level diagram contributing to the
effective Lagrangian. The solid lines represent lepton doublets and the
thick dashed line represents the heavy scalar. Diagram b) represents the
four-lepton interaction in the effective theory. The symbol $\otimes$ means
a scalar current insertion.

\vskip 0.4cm\noindent
{\bf Figure 2:}  Diagrammatic representation of the contributions to the
one-loop effective Lagrangian given by the determinant of the
fluctuation operator.  Gauge bosons are represented by wavy lines
and the standard Higgs is represented by  thin dashed lines.

\vskip 0.4cm\noindent
{\bf Figure 3:} Matching conditions: Self-energy diagrams. a) in the full
theory and b) in the effective one.
\vskip 0.4cm\noindent
{\bf Figure 4:}  Matching conditions: Penguin diagrams a) and b) in the
full theory and c) in the effective one.
\vskip 0.4cm\noindent
{\bf Figure 5:}  Matching conditions: Seagull diagrams a) and b) in the full
theory and c) in the effective one.
\vskip 0.4cm\noindent
{\bf Figure 6:}  Matching conditions: Box diagrams a) and b) in the full
theory and c) in the effective one.
\vskip 0.4cm\noindent
{\bf Figure 7:} The family lepton-number violating branching ratio
 $BR(Z\rightarrow \tau^- e^+)=
 \Gamma(Z\rightarrow \tau^- e^+)/\Gamma(Z\rightarrow e^- e^+)$
 computed in both
the full (solid line) and the effective (dashed line) theories as a
function of the charged scalar mass. The couplings $F_{\tau e}$ are
taken to be equal to 1.

\end{document}